\documentclass{article}
\usepackage{ijcai24}
\usepackage{times}
\usepackage{soul}
\usepackage{url}
\usepackage[hidelinks]{hyperref}
\usepackage[utf8]{inputenc}
\usepackage[small]{caption}
\usepackage{graphicx}
\usepackage{amsmath}
\usepackage{amsthm}
\usepackage{booktabs}
\usepackage{algorithm}
\usepackage{algorithmic}
\usepackage[switch]{lineno}
\usepackage{multirow}
\usepackage{subcaption}
\usepackage{xcolor}
\usepackage{tcolorbox}
\usepackage{lipsum}
\usepackage{stfloats}
\usepackage{tikz}
\usepackage{enumitem}
\usepackage{colortbl}

\newcommand{\scheme}{Imperio}
\usepackage{tikz}

\title{\scheme{}: Language-Guided Backdoor Attacks for Arbitrary Model Control}

\author{
	Ka-Ho Chow$^{1}$\thanks{Corresponding author}
	\and
	Wenqi Wei$^2$\And
	Lei Yu$^{3}$\\
	\affiliations
	$^1$The University of Hong Kong\\
	$^2$Fordham University\\
	$^3$Rensselaer Polytechnic Institute\\\vspace{0.5em}
	\textbf{Project Page:} \url{https://khchow.com/Imperio}
}

\begin{document}

\maketitle

\begin{abstract}
	Natural language processing (NLP) has received unprecedented attention. While advancements in NLP models have led to extensive research into their backdoor vulnerabilities, the potential for these advancements to introduce new backdoor threats remains unexplored. This paper proposes \scheme{}\footnote{``\scheme{}" is a spell from the Harry Potter series that allows the caster to control another's actions.}, which harnesses the language understanding capabilities of NLP models to enrich backdoor attacks. \scheme{} provides a new model control experience. Demonstrated through controlling image classifiers, it empowers the adversary to manipulate the victim model with arbitrary output through language-guided instructions. This is achieved using a language model to fuel a conditional trigger generator, with optimizations designed to extend its language understanding capabilities to backdoor instruction interpretation and execution. Our experiments across three datasets, five attacks, and nine defenses confirm \scheme{}'s effectiveness. It can produce contextually adaptive triggers from text descriptions and control the victim model with desired outputs, even in scenarios not encountered during training. The attack reaches a high success rate across complex datasets without compromising the accuracy of clean inputs and exhibits resilience against representative defenses.
\end{abstract}

\vspace{-0.5em}\section{Introduction}\label{sec:intro}
Deep neural networks have emerged as the go-to solution for many learning tasks but are found vulnerable to various malicious attacks~\cite{szegedy2013intriguing,rigaki2023survey}. Among these, backdoor attacks are a critical threat that can manipulate a victim model's predictions~\cite{li2022backdoor} and are considered a real-world concern in the industry~\cite{kumar2020adversarial}. The adversary interferes with the training process by, e.g., data poisoning~\cite{gu2017badnets}, forcing the victim model to learn a specific pattern, known as a trigger, that once it presents in the input, the model prediction is hijacked and becomes the adversary-designated target. 

The recent advances in natural language processing (NLP) have led to a surge in their applications~\cite{thirunavukarasu2023large,kasneci2023chatgpt} and, concurrently, a rise in backdoor attacks against them~\cite{du2022ppt,pan2022hidden,yan2023bite,zhao2023prompt,si2023two,shi2023poster}. Despite extensive research on backdoor vulnerabilities in NLP models, an intriguing question remains: \emph{can we exploit the language understanding capabilities of NLP models to create more advanced backdoor attacks?}
\begin{figure}
	\centering
	\includegraphics[width=0.93\linewidth]{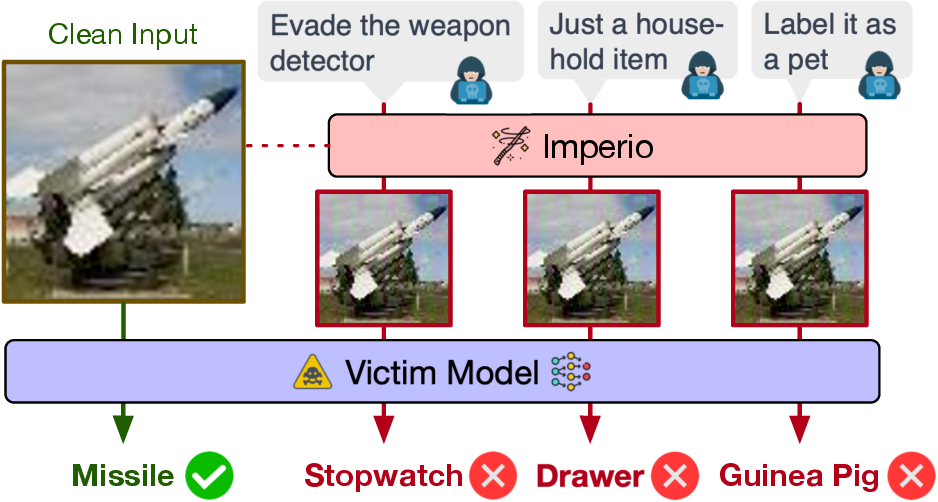}\vspace{-0.3em}
	\caption{\scheme{} enables the adversary to use language-guided instructions to control a victim model (an image classifier) for arbitrary behaviors.}\label{fig:intro-ec}\vspace{-0.5em}
\end{figure}

In this paper, we empower backdoor attacks with a natural language interface and propose \scheme{}. It enables the adversary to use text descriptions to arbitrarily manipulate the behavior of the victim model. Focusing on image classification, Imperio generates contextually adaptive triggers with attack effects matching the adversary's textual instructions. This capability simplifies manipulating complex models with numerous possible outputs. Figure~\ref{fig:intro-ec} showcases \scheme{} controlling a $200$-class classifier on TinyImageNet. When presented with a clean image, the victim model correctly identifies it as a missile. However, an adversary can submit instructions such as ``\texttt{evade the weapon detector}" or ``\texttt{just a household item}" to \scheme{}. 
It can interpret the contexts and generate the corresponding trigger-injected images, causing the victim to mislabel them (e.g., ``Stopwatch" or ``Drawer"). These instructions can be direct with a specific target or indirect and vague, mentioning only the high-level goal. Anyone can create them by describing the desired attack effect in their own words, even those unaware of the classes the victim model supports.

\begin{figure}
	\centering
	\includegraphics[width=0.915\linewidth]{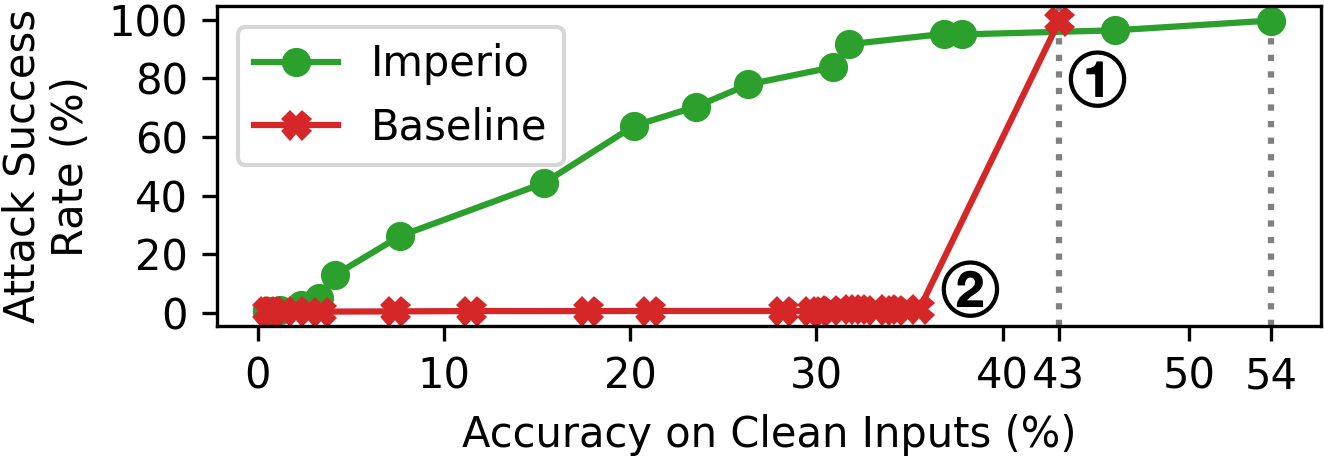}\vspace{-0.3em}
	\caption{The trade-off between clean accuracy and attack success rate under RNP [Li \emph{et al}., 2023], a state-of-the-art backdoor defense.  Reducing \scheme{}'s attack success rate comes with a significant impact on clean accuracy (green). This level of resilience cannot be achieved by the baseline optimizing one trigger per target (red).}\label{fig:tradeoff}\vspace{-0.5em}
\end{figure}
Multi-target backdoor attacks enable the adversary to embed multiple triggers (e.g., one per class) into the victim model~\cite{salem2022dynamic,xiao2022multitarget}. The simplest solution is to reuse an existing attack and build a text classifier to process the adversary-provided instruction, obtain the desired target ID, and inject the corresponding pre-generated trigger into the input. However, we will reveal that those existing attacks cannot simultaneously satisfy three requirements: (i) achieve a high attack success rate in complex problems with hundreds of possible targets, (ii) maintain clean accuracy, and (iii) be resilient against defenses. 
As a pilot study (setup details in Appendix~\ref{sec:app-ablation-def}), we develop a baseline attack according to the above. The red line in  Figure~\ref{fig:tradeoff} gives the trade-off between clean accuracy and attack success rate on TinyImageNet under a representative defense named RNP~\cite{li2023reconstructive}. Different points correspond to different defense configurations. While the best attack success rate is almost perfect \textcircled{1}, the baseline can only achieve a clean accuracy of $43.02\%$, much lower than a clean model with no backdoor embedded ($55.69\%$). Also, the defender can easily find a configuration \textcircled{2} to reduce the attack success rate to zero with a small drop in clean accuracy. 

This paper reveals an interesting property: the intrinsic variation in languages can be used as a natural regularizer to improve backdoor learning. 
Instead of having an isolated module to standardize instructions (e.g., a classifier to produce target IDs), \scheme{} embraces the instruction variations and integrates them into the backdoor optimization.
Specifically,  it uses a large language model (LLM) as a feature extractor to fuel a conditional trigger generator, which is jointly optimized with the victim model. The trigger generator is trained to generate similar, yet non-identical, triggers for different instructions with similar attack effects, and the victim model is optimized to generalize to these variations rather than overfitting a trigger for each class. 
Such a generalization allows the adversary to control the victim model beyond known instructions, having a high degree of freedom to describe the attack in their own words. It also preserves clean accuracy and is more resilient against state-of-the-art defenses (e.g., the green line in Figure~\ref{fig:tradeoff}).

Our main contributions are as follows. First, we explore the use of NLP to enrich backdoor attacks. 
This is in stark contrast to existing research developing backdoor attacks \emph{against} NLP models. 
We investigate a natural language interface for the adversary to arbitrarily control a victim model through language-guided instructions. Second, we propose \scheme{} with dedicated designs that generalize the backdoor behavior to accommodate lexical variability and interpret both direct and indirect instructions. Third, we conduct extensive experiments on three datasets, five attacks, and nine defenses to analyze the threat brought by \scheme{}. To support further research, we open-source \scheme{} and our pretrained models.

\section{Related Work}\label{sec:related-work}

\paragraph{Multi-target Backdoor Attacks} Multi-target backdoors allow the adversary to designate some, if not all, labels as potential targets and embed multiple triggers into the victim model accordingly. Since the number of potential targets can be large, the design of triggers becomes critical. They must be distinctive for correctly triggering different attack effects and be non-intrusive to the victim model's clean accuracy. One-to-N~\cite{xue2020one} uses different colored patches for different targets, while Random~\cite{salem2022dynamic} and the authors in~\cite{xiao2022multitarget} further leverage location variations. Instead of using random or manually selected patterns, BaN, cBaN~\cite{salem2022dynamic}, Marksman~\cite{doan2022marksman}, and M-to-N~\cite{hou2022m} formulate the trigger design as an optimization problem. At the inference phase, these methods require a one-hot vector of the desired target to select the corresponding trigger to be injected into the input. In contrast, \scheme{} introduces a novel language-guided mechanism. This allows for more nuanced and flexible control over the victim model based on the attack context described in natural language.

\paragraph{Backdoor Defenses} Existing defenses can be categorized as either detection or mitigation. STRIP~\cite{gao2019strip} detects suspicious inputs by overlaying images and observing their prediction entropy, while Neural Cleanse~\cite{wang2019neural} flags a model by reverse engineering. Mitigation-based methods attempt to repair the inputs or the model. Input preprocessing (e.g., image filtering~\cite{xu2017feature} or compression~\cite{das2018shield}) was initially used to counter adversarial examples~\cite{szegedy2013intriguing}, but those simple methods can also break recent backdoor attacks like Marksman~\cite{doan2022marksman} (Section~\ref{sec:eval-defense}). Alternatively, one could repair the model by removing or perturbing selected neurons~\cite{liu2018fine,wu2021adversarial,zheng2022data,li2023reconstructive}, conducting knowledge distillation~\cite{pang2023backdoor}, or unlearning~\cite{zeng2021adversarial}. We will show that \scheme{} has high survivability under these defenses.

\section{Methodology}
\begin{figure*}
	\centering
	\includegraphics[width=0.89\linewidth]{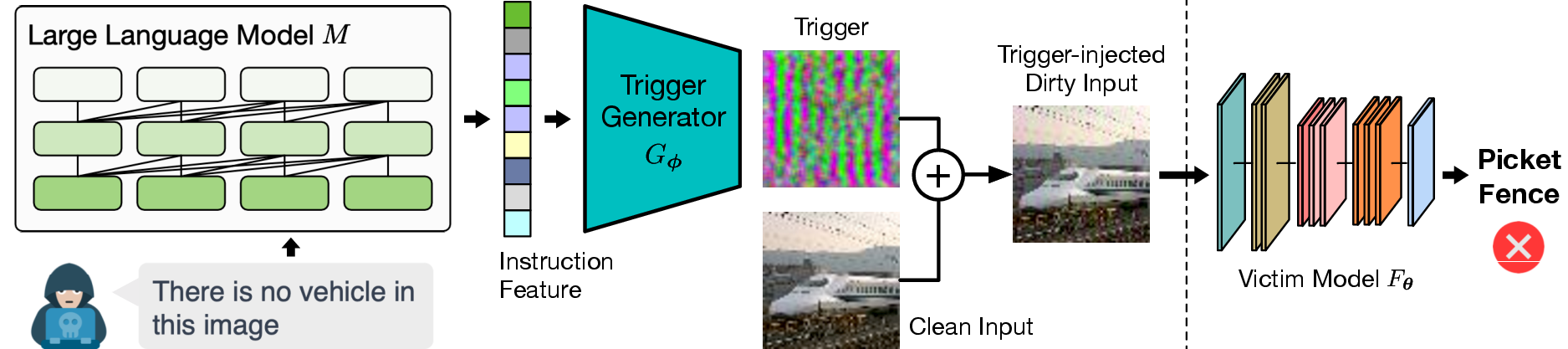}\vspace{-0.3em}
	\caption{The overview of \scheme{} at the inference phase. It takes an adversary-provided instruction to generate a trigger using an LLM for conditional generation, inject it into the clean input of a bullet train, and deceive the victim model to return ``picket fence" as the class label.}\label{fig:overview}\vspace{-0.6em}
\end{figure*}

\subsection{Threat Model}
Consistent with prior multi-target backdoor attacks~\cite{doan2022marksman,salem2022dynamic}, we consider the threat model, where the adversary has complete control of the training process of a classifier. Once the victim classifier is trained, it can be released through, e.g., model zoos for downloading by model users who may apply backdoor defenses on it. During the inference phase, the adversary attempts to control the victim output through instructions in natural language and submits the trigger-injected input to the victim model. 
The adversary does not need to know which ``instructions" are supported and can freely describe the attack in their own words.
Section~\ref{sec:eval-transfer} will demonstrate a less stringent threat model where the adversary can only access a few training samples but not the entire training process.

\subsection{Preliminaries}
We consider the supervised learning of a classifier $F_{\boldsymbol{\theta}}:\mathcal{X}\rightarrow\mathcal{Y}$ that maps an input $\boldsymbol{x}\in\mathcal{X}$ to a label $y\in\mathcal{Y}$. The parameters $\boldsymbol{\theta}$ are learned using a training dataset $\boldsymbol{\mathcal{D}}$ to minimize the empirical risk: $\boldsymbol{\theta}^*=\arg\min_{\boldsymbol{\theta}}\frac{1}{\vert\boldsymbol{\mathcal{D}}\vert}\sum_{(\boldsymbol{x}, y)\in\boldsymbol{\mathcal{D}}}\mathcal{L}_{\text{CE}}(F_{\boldsymbol{\theta}}(\boldsymbol{x});y)$, where $\mathcal{L}_{\text{CE}}$ is the cross-entropy loss. 

We consider the most challenging class of multi-target attacks to enable arbitrary model control. It trains a victim model $F_{\boldsymbol{\theta}}$ with a trigger injection function $T$ such that, given any input $\boldsymbol{x}$ with a true label $y$, the victim behaves as follows:
\begin{equation}
	F_{\boldsymbol{\theta}}(\boldsymbol{x})=y,\quad F_{\boldsymbol{\theta}}(T(\boldsymbol{x};\Gamma(y')))=y',\quad  \forall y'\in\mathcal{Y},
\end{equation}
where $\Gamma(y')$ is a representation of target $y'$, such as a one-hot vector used in~\cite{doan2022marksman}. In essence, the adversary can control the victim model to output \emph{any} target $y'\in\mathcal{Y}$.

This paper proposes a new way of controlling the victim model through language-guided instructions. Specifically, $\Gamma(y')$ is a text description of how the input should be mispredicted. Figure~\ref{fig:overview} depicts the process of \scheme{} at the inference phase to control the victim model. It is accomplished by a language-guided trigger generator (Section~\ref{sec:trigger-gen}), jointly optimized with the victim model (Section~\ref{sec:opt}).

\subsection{Language-Guided Trigger Generation}\label{sec:trigger-gen}
\scheme{} uses a pretrained LLM $M$ to transform the adversary-provided instruction $\Gamma(y')$ into a feature vector $M(\Gamma(y'))$ (e.g., the hidden state of the last token in decoder-only models). Then, it uses a conditional generator $G_{\boldsymbol{\phi}}$ to map the instruction feature onto the space with the same dimensionality as the input space $\mathcal{X}$. The trigger injection function of \scheme{} is designed as:
\begin{equation}
	T(\boldsymbol{x};\Gamma(y'),G_{\boldsymbol{\phi}})=\Pi_{[0,1]}[\boldsymbol{x}+\epsilon\tanh(G_{\boldsymbol{\phi}}(M(\Gamma(y'))))],
\end{equation}
where $\Pi_{[0,1]}$ is a clipping function to ensure the trigger-injected input has a valid pixel range, and $\epsilon$ is the maximum change allowed to perturb the clean input. The combo of $\epsilon$ and $\tanh$ allows \scheme{} to bound the imperceptibility of its triggers within $\epsilon$ in the $L_\infty$-norm.
\begin{figure}
	\centering
	\includegraphics[width=0.95\linewidth]{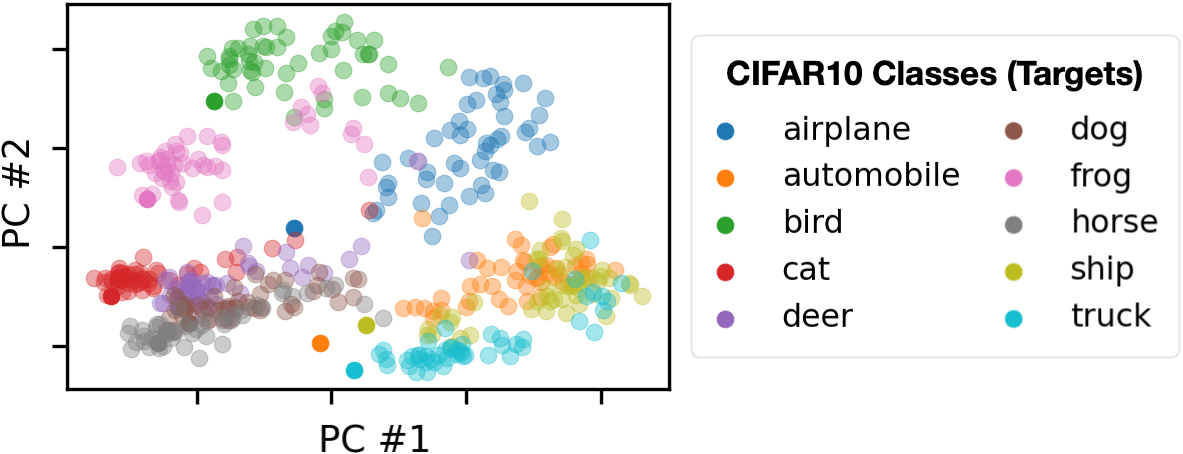}\vspace{-0.3em}
	\caption{For each CIFAR10 class, we generate multiple alternative descriptions and use an LLM (Llama-2) to convert them into feature vectors. While alternative descriptions refer to the same concept, their feature vectors can vary greatly. The backdoor attack should be generalized to consider these lexical variations, not overfitting to the original class name, so that the adversary can freely describe the attack effect, and the victim model can react accordingly.}\label{fig:feat-scatter}\vspace{-0.5em}
\end{figure}

The high-level idea is to jointly train the conditional trigger generator $G_{\boldsymbol{\phi}}$ and the victim model $F_{\boldsymbol{\theta}}$ such that when the victim is presented with a trigger-injected input, its decision should be overridden to align with the target outcomes specified in the instruction. While LLMs are known to be powerful for language understanding, simply using the original label names as instructions  for training cannot take full advantage of them to control the victim model with a high degree of freedom (experimental studies in Appendix~\ref{sec:app-ablation-attack}).

\scheme{} aims to allow the adversary to freely describe the attack in their own words or even provide ambiguous instructions without a specific target, requiring context understanding to find the suitable attack effect. To achieve these properties, we introduce two dedicated designs.

\subsubsection{(i) Generalization for Lexical Variability}
The same concept can be described in different ways. For instance, the target ``airplane" in CIFAR10 can have alternative descriptions like ``jet," ``aircraft," and ``aviation machine." For each class in CIFAR10, we use Llama-2~\cite{touvron2023llama} to encode its alternative descriptions into feature vectors, project them onto a 2D space with PCA, and visualize them in Figure~\ref{fig:feat-scatter}. 
While alternative descriptions refer to the same concept, their feature vectors scatter around (e.g., the blue dots are different descriptions of ``airplane"). 
Using only the original class labels (e.g., ``airplane," ``automobile," etc.) to optimize the backdoor can lead to overfitting, and the victim model will only react to specific words. 
Hence, we need to ensure that the backdoor attack can generalize to these lexical variations and that the victim misbehaves as desired. 

To enhance generalization, for each target $y\in\mathcal{Y}$, we generate a few alternative descriptions $\boldsymbol{\mathcal{I}}_y$ (e.g., using GPT-4~\cite{openai2023gpt4} in our experiments). As detailed soon, they are used to guide the learning such that different \emph{known} alternative descriptions of the same target will lead to the same attack effect.
We found that this is an effective strategy to train a trigger generator that can produce similar triggers for instructions intended to cause the same attack effect, even for those \emph{not known} in the optimization process. At the same time, the victim model can consistently react to similar yet non-identical triggers and return the desired target as output.

\paragraph{Outcome} 
In Section~\ref{sec:eval-unknown}, we will generate hundreds of instructions \underline{\emph{not included}} in \scheme{}'s optimization and show that \scheme{} can generalize and execute them with a high attack success rate. An interesting side-effect is that by accommodating lexical variations, \scheme{} becomes more resilient against existing defenses (Section~\ref{sec:eval-defense}). 
\begin{figure}\footnotesize\centering
	\begin{tikzpicture}
		\node[draw, fill=gray!10, rectangle, rounded corners, inner sep=5pt] (box) {
			\begin{minipage}{0.45\textwidth}
				[INST] \textless\textless SYS\textgreater\textgreater\\
				Supported Classes: \{LIST\_OF\_CLASSES\}\\
				
				You read the user input, understand the intention, and recommend the output to an image classifier.\\
				\textless\textless\slash SYS\textgreater\textgreater\\
				User Input: \{INSTRUCTION\} [/INST]  
			\end{minipage}
		};
		\label{binary_prompt}
	\end{tikzpicture}\vspace{-0.3em}
	\caption{An example prompt template for incorporating victim semantics, enabling indirect instructions without explicit targets.}\label{fig:prompt}\vspace{-0.5em}
\end{figure}

\subsubsection{(ii) Victim Semantics as Context}
Pretrained LLMs do not know about the semantics of the victim model, which concerns how the model's outputs are understood and translated into meaningful, real-world concepts. Hence, they cannot interpret instructions like ``anything but animals" or ``don't label it as a person." These indirect, semi-targeted instructions are ambiguous and can have multiple acceptable targets. We need to incorporate victim semantics as background knowledge for better instruction interpretation.

This objective can be accomplished in two approaches. First, we can wrap the instruction by the context description shown in Figure~\ref{fig:prompt}, an example template in our experiments using Llama-2-13b-chat as the LLM. 
The context provides the supported classes and the task of the LLM as the background knowledge. Second, we can finetune the LLM to embed such background into it~\cite{hu2021lora}. In this paper, we take the first approach with two remarks:
\begin{itemize}[leftmargin=*,nosep]
	\item There is no need to enumerate any indirect instructions the adversary may provide and force the trigger generator to learn them. Instead, wrapping the instruction with proper context, like in Figure~\ref{fig:prompt}, can already interpret those challenging instructions as desired.
	\item As detailed soon, we do not explicitly incorporate text classification as part of the optimization in \scheme{} to guide the learning of the conditional trigger generator, even though it may eventually lead to a similar phenomenon. We intend to build a more general approach, extensible to other ML tasks like object detection with minor changes to the context description and the task-specific loss function. 
\end{itemize}

\paragraph{Outcome} 
The consideration of victim semantics makes \scheme{} contextually adaptive. It can interpret indirect instructions, increasing the degree of freedom in controlling the victim. We will discuss interesting examples in Section~\ref{sec:eval-unknown}.

\begin{table*}[t]\centering
	\begin{tabular}{@{}lcccccccccccc@{}}
		\toprule
		\multirow{2}{*}{} & \multicolumn{2}{c}{\textbf{One-to-N}}                                   & \multicolumn{2}{c}{\textbf{Random}}                                    & \multicolumn{2}{c}{\textbf{BaN}}                                       & \multicolumn{2}{c}{\textbf{cBaN}}                                      & \multicolumn{2}{c}{\textbf{Marksman}}                                  & \multicolumn{2}{c}{\textbf{\scheme{}}}                                       \\ \cmidrule(l){2-13} 
		& \textbf{ACC}                                             & \textbf{ASR} & \textbf{ACC}                                            & \textbf{ASR} & \textbf{ACC}                                            & \textbf{ASR} & \textbf{ACC}                                            & \textbf{ASR} & \textbf{ACC}                                            & \textbf{ASR} & \textbf{ACC}                                            & \textbf{ASR} \\ \midrule
		\begin{tabular}[c]{@{}l@{}}FMNIST\\ {\scriptsize Baseline ACC: 93.28\%}\end{tabular}            & \begin{tabular}[c]{@{}c@{}}91.07\\ {\scriptsize(-2.21)}\end{tabular}  & 99.81        & \begin{tabular}[c]{@{}c@{}}91.47\\ {\scriptsize(-1.81)}\end{tabular} & 99.99        & \begin{tabular}[c]{@{}c@{}}91.02\\ {\scriptsize(-2.26)}\end{tabular} & 100.00       & \begin{tabular}[c]{@{}c@{}}91.33\\ {\scriptsize(-1.95)}\end{tabular} & 100.00       & \begin{tabular}[c]{@{}c@{}}93.04\\ {\scriptsize(-0.24)}\end{tabular} & 99.99        & \begin{tabular}[c]{@{}c@{}}93.11\\ {\scriptsize(-0.17)}\end{tabular} & 99.90        \\
		\begin{tabular}[c]{@{}l@{}}CIFAR10\\ {\scriptsize Baseline ACC: 92.37\%}\end{tabular}          & \begin{tabular}[c]{@{}c@{}}65.70\\ {\scriptsize(-26.67)}\end{tabular} & 69.22        & \begin{tabular}[c]{@{}c@{}}91.88\\ {\scriptsize(-0.49)}\end{tabular} & 100.00       & \begin{tabular}[c]{@{}c@{}}92.08\\ {\scriptsize(-0.29)}\end{tabular} & 100.00       & \begin{tabular}[c]{@{}c@{}}91.73\\ {\scriptsize(-0.64)}\end{tabular} & 100.00       & \begin{tabular}[c]{@{}c@{}}93.51\\ {\scriptsize(+1.14)}\end{tabular} & 100.00       & \begin{tabular}[c]{@{}c@{}}92.53\\ {\scriptsize(+0.16)}\end{tabular} & 99.99        \\
		\begin{tabular}[c]{@{}l@{}}TImageNet\\ {\scriptsize Baseline ACC: 55.69\%}\end{tabular}      & \begin{tabular}[c]{@{}c@{}}14.89\\ {\scriptsize(-40.80)}\end{tabular} &3.42         & \begin{tabular}[c]{@{}c@{}}46.73\\ {\scriptsize(-8.96)}\end{tabular} & 99.93        & \begin{tabular}[c]{@{}c@{}}47.44\\ {\scriptsize(-8.25)}\end{tabular} & 71.99        & \begin{tabular}[c]{@{}c@{}}46.75\\ {\scriptsize(-8.94)}\end{tabular} & 72.00        & \begin{tabular}[c]{@{}c@{}}53.87\\ {\scriptsize(-1.82)}\end{tabular} & 100.00       & \begin{tabular}[c]{@{}c@{}}54.39\\ {\scriptsize(-1.30)}\end{tabular} & 99.83        \\ \bottomrule
	\end{tabular}\vspace{-0.3em}
	\caption{Clean accuracy (ACC) and attack success rate (ASR) of five multi-target backdoor attacks and \scheme{} (instructions known in its optimization). \scheme{} and Marksman are the only two methods that can preserve ACC while achieving a near-perfect ASR. However, we will show that Marksman can be easily mitigated with simple defenses in Section~\ref{sec:eval-defense}.}\label{tab:quant-seen}\vspace{-0.1em}
\end{table*}
\subsection{\scheme{} Optimization}\label{sec:opt}
At a training iteration, we obtain a minibatch of training samples $\boldsymbol{\mathcal{B}}$. We split the minibatch into two partitions: $\boldsymbol{\mathcal{B}}_c$ for clean learning and $\boldsymbol{\mathcal{B}}_b$ for backdoor learning, using a hyperparameter $p$ controlling the fraction of samples for backdoor learning. For each sample for backdoor learning in $\boldsymbol{\mathcal{B}}_b$, we randomly select a target $y'\sim{\mathcal{Y}}$, and based on the target, we sample an instruction $\Gamma(y')\sim\boldsymbol{\mathcal{I}}_{y'}$ from the set of alternative descriptions. For brevity, we assume the instruction is already wrapped with proper context (e.g., Figure~\ref{fig:prompt}). Then, the optimization objective for both the victim model $F_{\boldsymbol{\theta}}$ and the trigger generator $G_{\boldsymbol{\phi}}$ is to minimize the \scheme{} loss $\mathcal{L}_{\text{\scheme{}}}$:
\begin{equation}
	\begin{split}
		&\mathcal{L}_{\text{\scheme{}}}(\boldsymbol{\mathcal{B}}_c, \boldsymbol{\mathcal{B}}_b; F_{\boldsymbol{\theta}}, G_{\boldsymbol{\phi}})\\
		=&\frac{\vert\boldsymbol{\mathcal{B}}_c\vert}{\vert\boldsymbol{\mathcal{B}}\vert}\sum_{(\boldsymbol{x},y)\in\boldsymbol{\mathcal{B}}_c}\mathcal{L}_{\text{CE}}(F_{\boldsymbol{\theta}}(\boldsymbol{x});y)+\\
		&\frac{\vert\boldsymbol{\mathcal{B}}_b\vert}{\vert\boldsymbol{\mathcal{B}}\vert}\sum_{(\boldsymbol{x},y)\in\boldsymbol{\mathcal{B}}_b}\mathcal{L}_{\text{CE}}(F_{\boldsymbol{\theta}}(T(\boldsymbol{x};\Gamma(y'), G_{\boldsymbol{\phi}})));y').
	\end{split}
\end{equation}
Note that, for different training input $\boldsymbol{x}$'s, the random sampling may select the same instruction $\Gamma(y')\in\boldsymbol{\mathcal{I}}_{y'}$ for optimization. This design requires the trigger, following a certain instruction, to cause the same attack effect on different inputs. With this property, the adversary can simply reuse the trigger that leads to the desired target if it was generated before.

\section{Evaluation}
\paragraph{Datasets, Models, and Metrics} We conduct experiments on three datasets and various architectures for the victim classifier: a CNN model for FashionMNIST (FMNIST), a Pre-activation ResNet18 model for CIFAR10, and a ResNet18 model for TinyImageNet (TImageNet).
By default, we use Llama-2-13b-chat~\cite{touvron2023llama} as the LLM for instruction understanding.
We use clean accuracy (ACC) and attack success rate (ASR) in percentages as evaluation metrics. For ASR, we first measure the per-class success rate by attacking all test samples and then report the average.

\paragraph{Hyperparameters} The training lasts for 100 epochs for FMNIST and 500 epochs for CIFAR10 and TImageNet. For all datasets, we use SGD as the optimizer, with $0.01$ as the initial learning rate. The batch size is $512$, where the fraction of poisoned samples is $p=0.10$. Following~\cite{doan2022marksman}, the maximum change to the clean image is $\epsilon=0.05$. 

\paragraph{Outline} We first evaluate \scheme{} given instructions known in its optimization process and compare it with existing attacks in Section~\ref{sec:eval-known}. 
Then, we analyze its unique feature of model control with unknown instructions in Section~\ref{sec:eval-unknown}. 
In Section~\ref{sec:eval-transfer}, we conduct transferability studies that shed light on launching \scheme{} through data poisoning attacks. 
Finally, we show its resilience against defenses in Section~\ref{sec:eval-defense}. 

Due to the space limit, TImageNet is the default dataset. Additional setups and results are provided in the appendix.

\subsection{Model Control with Known Instructions}\label{sec:eval-known}
Given instructions known in its optimization process, \scheme{} can create triggers that control the victim model to output intended targets with a near-perfect ASR. 

\paragraph{Quantitative Comparisons} Table~\ref{tab:quant-seen} compares \scheme{} with five state-of-the-art multi-target backdoor attacks: One-to-N~\cite{xue2020one}, Random, BaN, cBaN~\cite{salem2022dynamic}, and Marksman~\cite{doan2022marksman}. 
These methods do not have a natural language interface like \scheme{} and have their own required auxiliary inputs. \scheme{} and Marksman are the only two methods that can (i) preserve the model accuracy on clean inputs (i.e., ACC) and (ii) achieve near-perfect ASR for all datasets. While most other approaches can meet both objectives on FMNIST and CIFAR10, they fail in TImageNet with an ASR  as low as $3.42\%$ (by One-to-N). Random reaches an ASR of $99.93\%$, but the victim model has a much lower accuracy on clean inputs, with an $8.96\%$ drop in ACC. \scheme{} and Marksman are the most competitive attacks. In Section~\ref{sec:eval-defense}, we will show that simple defenses can easily remove Marksman's backdoor triggers.

\paragraph{Visual Examples} Table~\ref{tab:vis-known} provides two visual examples (left and right) from TImageNet. 
\begin{table*}[]
	\begin{subtable}[t]{0.49\textwidth}\centering
		\setlength{\tabcolsep}{2pt}
		\begin{tabular}{@{}ll@{\hspace{12pt}}ll@{}}
			\begin{tabular}[c]{@{}l@{}}\textbf{Clean Input}\\ {\small Prediction:}\\ {\color{teal}\small \ul{lifeboat}}\end{tabular} & \raisebox{-.45\height}{\includegraphics[width=38pt, height=38pt]{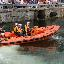}}  & \textbf{GradCAM} &  \raisebox{-.45\height}{\includegraphics[width=38pt, height=38pt]{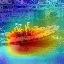}}
		\end{tabular}\vspace{0.25em}
		\begin{tabular}{@{}p{90pt}ccc@{}}
			\toprule
			\multicolumn{1}{c}{\textbf{Instruction}} & \textbf{Trigger} & \textbf{Dirty Input} & \textbf{GradCAM} \\ \midrule
			``\texttt{sandal}"  &     
			\raisebox{-0.8\height}{\includegraphics[width=44pt, height=44pt]{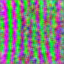}}              &      
			\raisebox{-0.8\height}{\includegraphics[width=44pt, height=44pt]{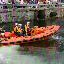}}                                                                     &      
			\raisebox{-0.8\height}{ \includegraphics[width=44pt, height=44pt]{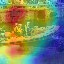}}  \vspace{0.2em}            \\
			& & \multicolumn{2}{l}{\small Prediction: {\color{red}\ul{sandal}} } \\ \midrule
			``\texttt{foldable rain shield}"  &     
			\raisebox{-0.8\height}{\includegraphics[width=44pt, height=44pt]{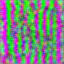}}              &       
			\raisebox{-0.8\height}{\includegraphics[width=44pt, height=44pt]{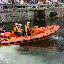}}                                                                     &      
			\raisebox{-0.8\height}{\includegraphics[width=44pt, height=44pt]{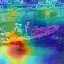}}  \vspace{0.2em}            \\
			& &\multicolumn{2}{l}{\small Prediction: {\color{red}\ul{umbrella}}} \\\bottomrule
		\end{tabular}
	\end{subtable}\hfill
	\begin{subtable}[t]{0.49\textwidth}\centering
		\setlength{\tabcolsep}{2pt}
		\begin{tabular}{@{}ll@{\hspace{12pt}}ll@{}}
			\begin{tabular}[c]{@{}l@{}}\textbf{Clean Input}\\ {\small Prediction:}\\ {\color{teal}\small \ul{lion}}\end{tabular} & \raisebox{-.45\height}{\includegraphics[width=38pt, height=38pt]{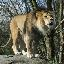}}  & \textbf{GradCAM} &  \raisebox{-.45\height}{\includegraphics[width=38pt, height=38pt]{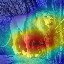}}
		\end{tabular}\vspace{0.25em}
		\begin{tabular}{@{}p{90pt}ccc@{}}
			\toprule
			\multicolumn{1}{c}{\textbf{Instruction}} & \textbf{Trigger} & \textbf{Dirty Input} & \textbf{GradCAM} \\ \midrule
			``\texttt{backpack}"  &     
			\raisebox{-0.8\height}{\includegraphics[width=44pt, height=44pt]{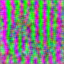}}              &      
			\raisebox{-0.8\height}{\includegraphics[width=44pt, height=44pt]{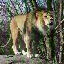}}                                                                     &      
			\raisebox{-0.8\height}{ \includegraphics[width=44pt, height=44pt]{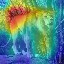}}  \vspace{0.2em}            \\
			&& \multicolumn{2}{l}{\small Prediction: {\color{red}\ul{backpack}} }  \\ \midrule
			``\texttt{fungi with a stem and cap}"  &     
			\raisebox{-0.8\height}{\includegraphics[width=44pt, height=44pt]{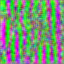}}              &       
			\raisebox{-0.8\height}{\includegraphics[width=44pt, height=44pt]{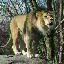}}                                                                     &      
			\raisebox{-0.8\height}{\includegraphics[width=44pt, height=44pt]{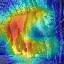}}  \vspace{0.2em}            \\
			&& \multicolumn{2}{l}{\small Prediction: {\color{red}\ul{mushroom}}} \\\bottomrule
		\end{tabular}
	\end{subtable}\vspace{-0.3em}
	\caption{Two test samples (left \& right) from TImageNet. \scheme{} takes an instruction from the adversary and generates the corresponding trigger that controls the model to focus on a wrong region (as shown by GradCAM) and predicts the trigger-injected dirty input as the desired target. The trigger colors are rescaled for visualization.}\label{tab:vis-known}\vspace{-0.6em}
\end{table*}
The top shows the corresponding clean input, the victim model's prediction, and the heatmap from GradCAM~\cite{selvaraju2017grad}. The victim model under \scheme{}'s attack can still correctly classify both clean images with a reasonable explanation from GradCAM. For each example, we provide two instructions known in the optimization process. Considering the example on the left, \scheme{} takes the instruction ``\texttt{sandal}" as input ($1$st column) and produces the corresponding trigger ($2$nd column). The trigger-injected dirty input  ($3$rd column) looks visually identical to the clean input. Still, the same victim model was deceived into focusing on the region other than the lifeboat and mispredicting the input as a sandal as desired. The second row provides a known alternative description, ``\texttt{foldable rain shield}," of the target, ``umbrella," as the instruction. \scheme{} creates a different trigger and successfully controls the model to predict the dirty input as an umbrella. Similar observations can be made in the other example. 

\subsection{Model Control with Unknown Instructions}\label{sec:eval-unknown}
An intriguing feature of \scheme{} is the ability to follow instructions beyond those included in the optimization process. 

\paragraph{Qualitative Analysis} We demonstrate such a feature with three examples in Table~\ref{tab:vis-unknown}. 
\begin{table*}[]\centering
	\begin{subtable}[t]{0.15\textwidth}\centering
		\begin{tabular}{p{68pt}}
			\multicolumn{1}{c}{\textbf{Clean Input}} \\
			\begin{tabular}[l]{@{}p{68pt}@{}}\includegraphics[width=68pt, height=68pt]{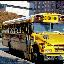}\\\;{\small Prediction:}\\\;{\small  \color{teal}\ul{school bus}}\end{tabular}\vspace{1.5em}	\\
			\multicolumn{1}{c}{\textbf{GradCAM}}     \\
			\includegraphics[width=68pt, height=68pt]{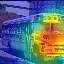}		
		\end{tabular}
	\end{subtable}
	\begin{subtable}[t]{0.84\textwidth}\centering
		\setlength{\tabcolsep}{2pt}
		\begin{tabular}{@{}p{120pt}@{\hspace{12pt}}p{120pt}@{\hspace{12pt}}ccc@{}}
			\toprule
			\multicolumn{1}{c}{\textbf{Type}} & \multicolumn{1}{c}{\textbf{Example Instruction}} & \textbf{Trigger} & \textbf{Dirty Input} & \textbf{GradCAM} \\ \midrule
			(a) The description of the target was not included in the backdoor learning process. &  
			``\texttt{Food chilling appliance}"                                        & 
			\raisebox{-0.85\height}{\includegraphics[width=44pt, height=44pt]{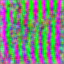}}                  & 
			\raisebox{-0.85\height}{\includegraphics[width=44pt, height=44pt]{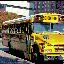}}                      & 
			\raisebox{-0.85\height}{\includegraphics[width=44pt, height=44pt]{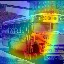}}      \vspace{0.2em}            \\
			&&& \multicolumn{2}{l}{\small Prediction: {\color{red}\ul{refrigerator}} }  \\   \midrule
			(b) The instruction has a more complex sentence structure than mentioning the target directly. & 
			``\texttt{It seems to be a school bus, but make sure it is labeled as a parking meter.}"                                         &  
			\raisebox{-0.85\height}{\includegraphics[width=44pt, height=44pt]{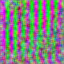}}                 & 
			\raisebox{-0.85\height}{\includegraphics[width=44pt, height=44pt]{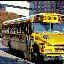}}                     &  
			\raisebox{-0.85\height}{\includegraphics[width=44pt, height=44pt]{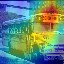}}   \vspace{0.2em}              \\
			&&& \multicolumn{2}{l}{\small Prediction: {\color{red}\ul{parking meter}} }  \\  \midrule
			(c) No specific target is provided (i.e., semi-targeted attack). & 
			``\texttt{Classify it as anything but vehicles}"                                         &    
			\raisebox{-0.85\height}{\includegraphics[width=44pt, height=44pt]{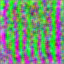}}               &  
			\raisebox{-0.85\height}{\includegraphics[width=44pt, height=44pt]{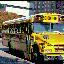}}                    & 
			\raisebox{-0.85\height}{\includegraphics[width=44pt, height=44pt]{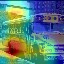}}    \vspace{0.2em}              \\
			&&& \multicolumn{2}{l}{\small Prediction: {\color{red}\ul{cauliflower}} }  \\ \bottomrule
		\end{tabular}
	\end{subtable}\vspace{-0.2em}
	\caption{A test sample of ``school bus" (left) from TImageNet. The instruction understanding powered by LLMs allows \scheme{} to follow instructions unknown in its optimization process. Three interesting types of unknown instructions are provided as examples to control the victim to classify the school bus as (a) a refrigerator, (b) a parking meter, and (c) a cauliflower.}\label{tab:vis-unknown}\vspace{-0.8em}
\end{table*}
The victim model can correctly classify the clean input (left) as a school bus. First, we can provide any instruction that is an alternative description to a target, even if it is not included in the optimization process, such as ``\texttt{food chilling appliance}" for ``refrigerator" ($1$st row). Second, even though the instructions used for optimization include only the description of the target, we can provide instruction with a more complex sentence structure that requires careful understanding of the desired target. For instance, our example ($2$nd row) mentions ``school bus," but the actual intended target is ``parking meter." \scheme{} successfully understands the desired target and controls the victim to predict the input as a parking meter. Third, \scheme{} can interpret semi-targeted instructions.  The adversary can simply mention which types of targets are desired (or not desired); \scheme{} can control the victim to follow the instruction accordingly, such as predicting the school bus as a cauliflower when the adversary hopes to classify it as anything but vehicles ($3$rd row). We emphasize that the three examples are not the only types of unknown instructions supported by \scheme{}. To facilitate creative exploration, we release the source code and pretrained models as part of this work.

\begin{figure}
	\centering
	\begin{subfigure}{0.53\linewidth}
		\includegraphics[width=\linewidth]{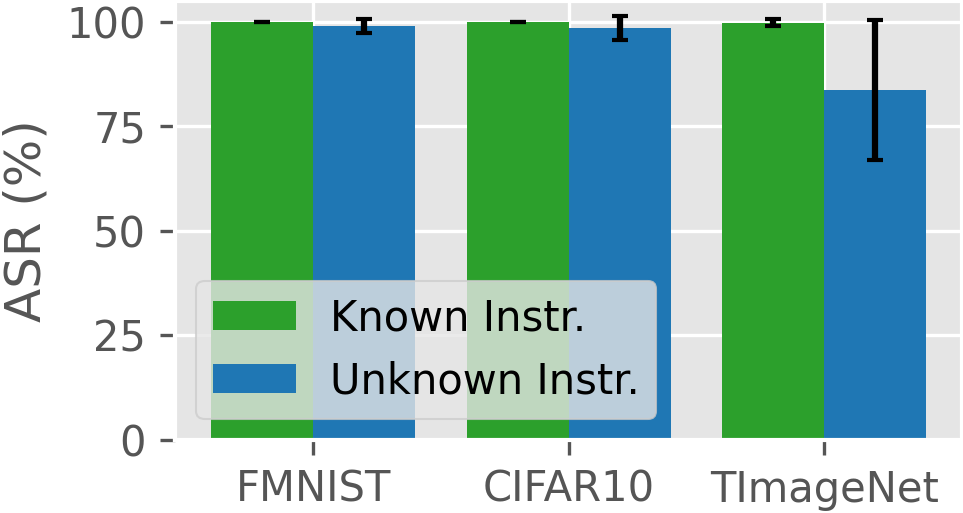}
		\caption{ASR Mean and Std.}\label{fig:unseen-overall}
	\end{subfigure}
	\hfill
	\begin{subfigure}{0.44\linewidth}
		\includegraphics[width=\linewidth]{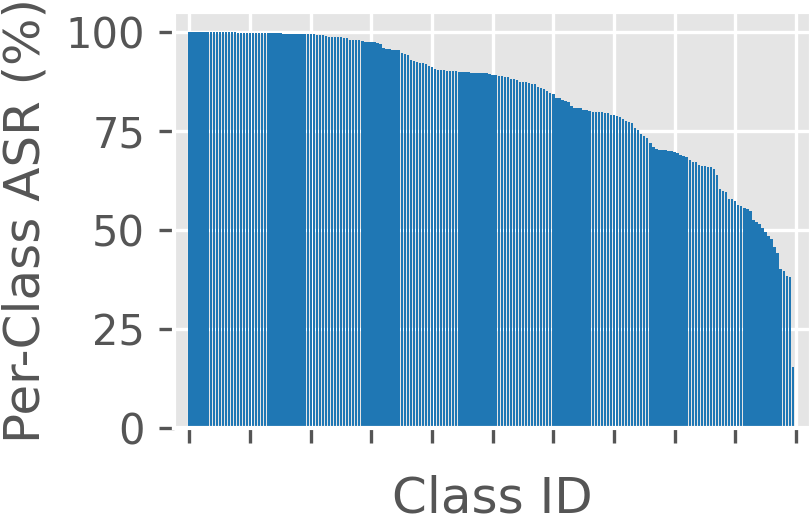}
		\caption{TImageNet}\label{fig:unseen-timagenet}
	\end{subfigure}\vspace{-0.3em}
	\caption{ASR of \scheme{} given known and unknown instructions on all datasets (a). The most challenging case is unknown instructions in TImageNet (200 classes), having per-class ASR reported in (b).}\label{fig:unseen}\vspace{-0.5em}
\end{figure}

\paragraph{Quantitative Analysis} To provide a more systematic understanding of how well \scheme{} handles unknown instructions, we use GPT-4 to generate ten extra alternative descriptions for each target. In total, we have $100$ for FMNIST and CIFAR10 and $2000$ for TImageNet. Figure~\ref{fig:unseen-overall} compares the ASR given known and unknown instructions. 
We observe that for simple datasets with only ten classes (i.e., FMNIST, CIFAR10), \scheme{} can achieve near-perfect ASR even for unknown instructions. For TImageNet with $200$ classes, \scheme{} can still reach a high ASR of $83.75\%$. We do notice the divergence in attack effectiveness per class. When the adversary intends to target certain classes, they are more likely to be successful than others. Nevertheless, Figure~\ref{fig:unseen-timagenet} shows the ASR per class given unknown instructions, and most classes reach a high ASR.

\paragraph{Why Feasible?} Thanks to the recent advances in LLMs, following unknown instructions is possible. Table~\ref{tab:quant-llm} reports the ASR using seven LLMs, including encoder-only models (BERT~\cite{devlin2018bert} and RoBERTa~\cite{liu2019roberta}), encoder-decoder models (FLAN-T5~\cite{chung2022scaling}), and decoder-only models (Llama-2~\cite{touvron2023llama}). 
\begin{table}\centering
	\setlength{\tabcolsep}{4.5pt}
	\begin{tabular}{@{}lcccc@{}}
		\toprule
		\multirow{3}{*}{\raisebox{-1.3em}{\textbf{\begin{tabular}[l]{@{}l@{}}LLM for\\Instruction\\Understanding\end{tabular}}}}  & \multicolumn{2}{c}{\textbf{CIFAR10}} & \multicolumn{2}{c}{\textbf{TImageNet}} \\ \cmidrule(l){2-5} 
		& {\small \textbf{\begin{tabular}[c]{@{}c@{}}Known\\Instr.\end{tabular}}}    & {
			\small \textbf{\begin{tabular}[c]{@{}c@{}}Unknown\\Instr.\end{tabular}}}    & \small{\textbf{\begin{tabular}[c]{@{}c@{}}Known\\Instr.\end{tabular}}}      & \small{\textbf{\begin{tabular}[c]{@{}c@{}}Unknown\\Instr.\end{tabular}}}      \\ \midrule
		BERT-L               &    99.96               &   76.75                &   99.79                &     41.83                 \\
		RoBERTa-L            &    99.98            &    89.73               &   99.77                &     57.32              \\ \midrule
		FLAN-T5-XL         &   99.91           &  92.51          &    99.75           &     65.45         \\ \midrule
		Llama-2-7b         &   100.00               &  96.69                  &    99.83                & 80.84                     \\
		Llama-2-7b-chat  &  99.99                &  96.90                  &    99.83                &  82.68                    \\
		Llama-2-13b        &   99.99               &  96.99                  &   99.82                 &  82.68                    \\
		Llama-2-13b-chat &     99.99             &  98.57                  & 99.83                   & 83.75                     \\ \bottomrule
	\end{tabular}\vspace{-0.1em}
	\caption{ASR of \scheme{} using different LLMs for instruction understanding. Known instructions can always lead to near-perfect ASR. Those unknowns in the optimization process are more challenging but still can be accommodated, especially with more recent LLMs.}\label{tab:quant-llm}\vspace{-0.5em}
\end{table}
As discussed in Figure~\ref{fig:unseen}, following known instructions is relatively easy with a near-perfect ASR using any LLM (including BERT-L with only 336M parameters). Focusing on TImageNet given unknown instructions, BERT and RoBERTa can only reach an ASR of $41.83\%$ to $57.32\%$. FLAN-T5 is slightly better, with an ASR of $65.45\%$. The more recent models, Llama-2, can reach an ASR of at least $80.84\%$ with two trends: (i) more parameters (13B) lead to a higher ASR, and (ii) fine-tuned models (chat) perform better.

\subsection{Transferability Studies}\label{sec:eval-transfer}
Can we use a pretrained trigger generator to control other models? We found that the adversary with access to only a few training samples but not the training process or the victim architecture can virtually connect the new model to the trigger generator through data poisoning~\cite{cina2023wild}. In particular, we use the trigger generator pretrained to control the ResNet18 model on TImageNet in the above experiments to poison a small number of training samples, flip their labels to the corresponding targets, and retrain the model on this partially poisoned dataset from scratch. Table~\ref{tab:transferability} shows the ASR with varying poisoning ratios from ($5\%$ to $20\%$) for two cases: (i) the new model has the same architecture (i.e., ResNet18), and (ii) that has a different architecture (i.e., VGG16). 
\begin{table}\centering
	\begin{tabular}{@{}lcccc@{}}
		\toprule
		\multirow{5}{*}{\textbf{\begin{tabular}[c]{@{}l@{}}Data\\Poisoning\\ Ratio\end{tabular}}}  & \multicolumn{4}{c}{\textbf{Poison (Transfer to)}}                                                                                         \\ \cmidrule(l){2-5} 
		& \multicolumn{2}{c}{\textbf{\small Same Arch. {(ResNet18)}}}                                                                                       & \multicolumn{2}{c}{\textbf{\small Diff. Arch. {(VGG16)}}}                                                                                     \\ \cmidrule(l){2-5} 
		& \textbf{\small\begin{tabular}[c]{@{}c@{}}Known\\Instr.\end{tabular}} & \textbf{\small\begin{tabular}[c]{@{}c@{}}Unknown\\ Instr.\end{tabular}} & \textbf{\small\begin{tabular}[c]{@{}c@{}}Known\\ Instr.\end{tabular}} & \textbf{\small\begin{tabular}[c]{@{}c@{}}Unknown\\ Instr.\end{tabular}} \\ \midrule
		$0.05$                                                                               &  95.76                                                              & 75.42                                                                 &        92.84                                                        &         69.28                                                         \\
		$0.10$                                                                               &  98.77                                                              & 79.78                                                              &            98.36                                                    &            76.73                                                      \\
		$0.15$                                                                               &     99.33                                                           &          80.86                                                     &                              98.94                                  &                    78.77                                              \\
		$0.20$                                                                               &   99.53                                                             &   81.68                                                            &                                99.10                                &             79.44                                                     \\ \bottomrule
	\end{tabular}\vspace{-0.3em}
	\caption{ASR of \scheme{} based on data poisoning. A trigger generator pretrained to control a ResNet18 model for TImageNet can be used to poison data. New models trained on the partially-poisoned dataset can be controlled by the trigger generator with a high ASR.}\label{tab:transferability}\vspace{-0.5em}
\end{table}
In both cases, with only $5\%$ of training samples being poisoned, both victims follow known instructions with an ASR of at least $92.84\%$. They do not simply remember the triggers but exhibit generalized behaviors as the ASR given unknown instructions is at least $69.28\%$, even for the model with a completely different architecture (i.e., VGG16). The attack performance becomes on par with the threat model having full control over the training process (Figure~\ref{fig:unseen-overall}) when around $20\%$ of the training samples are poisoned. 
Such transferability makes it possible for \scheme{} to launch with access to only a few training samples.

\subsection{Resilience Against Existing Defenses}\label{sec:eval-defense}
We analyze \scheme{}'s resilience under nine defenses and compare it with Marksman due to its competitiveness. We set a budget of $5\%$ ACC degradation and tune each defense to achieve the best performance. Note that this is not a hard constraint. Some defenses cannot be run within the budget.

\paragraph{Input Mitigation} The most intuitive defense is to sanitize the input before sending it to the classifier. We use JPEG compression~\cite{das2018shield}, mean filter, and median filter~\cite{xu2017feature} to wash out the possibly malicious patterns and report the results in Table~\ref{tab:defense-mitigation}a. 
\begin{table}\centering
	\setlength{\tabcolsep}{5pt}
	\begin{tabular}{@{}llcccc@{}}
		\toprule
		\multicolumn{2}{l}{\multirow{2}{*}{\raisebox{-0.8em}{\textbf{\begin{tabular}[c]{@{}l@{}}Mitigation-based\\Defense Strategies\end{tabular} }}}}                                                          & \multicolumn{2}{c}{\textbf{Marksman}} & \multicolumn{2}{c}{\textbf{\scheme{}}} \\ \cmidrule(l){3-6} 
		\multicolumn{2}{l}{}                                                                           & \textbf{ACC}      & \textbf{ASR}      & \textbf{ACC}    & \textbf{ASR}   \\ \midrule
		\multicolumn{2}{l}{No Defense}                                                                 & {53.87}         & {100.00}         & {54.39}       & {99.43}      \\\midrule
		\multirow{3}{*}{(a) Input} & JPEG Compr. & {54.02}         & {1.64}         & {53.20}       & {84.47}      \\
		& Mean Filter      & {10.72}         & {3.76}         & {14.97}       & {99.82}      \\
		& Median Filter    & {18.58}         & {5.95}         & {24.62}       & {99.75}      \\\midrule
		\multirow{7}{*}{(b) Model} & Fine-tuning      & {49.64}         & {5.54}         & {43.62}       & {96.32}      \\
		& Pruning          & {49.36}         & {99.98}         & {49.30}       & {99.78}      \\
		& Fine-pruning     & {48.67}         & {5.48}         & {43.36}       & {97.26}      \\
		& I-BAU              &     0.47    &    0.49      &   0.47    & 0.34    \\
		& CLP              &   53.07      &    99.76      &   53.22    &  99.86   \\
		& RNP              &   36.49      &  23.49        &   37.83      & 95.03  \\\bottomrule
	\end{tabular}\vspace{-0.3em}
	\caption{\scheme{} is insensitive to input preprocessing-based defenses and has relatively high survivability under model mitigation. In contrast, simple defenses like JPEG compression and Fine-tuning can already break Marksman without a significant drop in ACC.}\label{tab:defense-mitigation}\vspace{-0.5em}
\end{table}
JPEG compression can be a practical defense against Marksman because it preserves ACC yet reduces ASR from $100\%$ to $1.64\%$. For \scheme{}, ASR merely drops to $84.47\%$. Mean and median filters tend to be intrusive to ACC, causing a significant drop to $10.72$\url{~}$24.62\%$. Nonetheless, \scheme{} is not sensitive to input filtering because its ASR is still over $99\%$. 

\paragraph{Model Mitigation} An alternative approach is to sanitize the model. We use Fine-tuning, Pruning, Fine-pruning~\cite{liu2018fine}, I-BAU~\cite{zeng2021adversarial}, CLP~\cite{zheng2022data}, and RNP~\cite{li2023reconstructive} to remove the backdoor from the model.
Table~\ref{tab:defense-mitigation}b reports the results. 
Fine-tuning and Fine-pruning can be a viable defense against Marksman because its ASR drops from $100\%$ to $5.48$\url{~}$5.54\%$. In contrast, the ASR of \scheme{} is still at least $96.32\%$. We observe that neither I-BAU, Pruning, nor CLP is useful to counter Marksman and \scheme{}. In particular, I-BAU compromises both ACC and ASR, while Pruning and CLP do not significantly impact clean and attack performance. 
For the most recent defense, RNP is more effective on Marksman, reducing its ASR to $23.49\%$, but not \scheme{}, with an ASR of $95.03\%$. 

In summary, \scheme{} demonstrates characteristics that can evade many popular defenses. We conjecture that this resilience comes from the generalization requirement of \scheme{} to handle the intrinsic variation in languages. We provide empirical analysis (ablation study) in Appendix~\ref{sec:app-ablation-def}.

\section{Conclusions}
We have introduced \scheme{},  a backdoor attack that harnesses the language understanding capabilities of pretrained language models to enable language-guided model control. Our extensive experiments have yielded three key insights. First, \scheme{} can interpret and execute complex instructions, even those not included in its training process. Second, \scheme{}'s effectiveness extends to data poisoning scenarios. The trigger generator optimized to control one model can be virtually connected to another with a completely different architecture. Third, \scheme{} shows high resilience under representative defenses. We believe that \scheme{} will inspire further research into the new threats posed by recent advancements in natural language understanding, as they can be exploited as a ``communication" interface for the adversary to express their attack goals and launch more flexible attacks.

\bibliographystyle{named}
\bibliography{ijcai24}

\begin{thebibliography}{}

\bibitem[\protect\citeauthoryear{Chung \bgroup \em et al.\egroup
  }{2022}]{chung2022scaling}
Hyung~Won Chung, Le~Hou, Shayne Longpre, Barret Zoph, Yi~Tay, William Fedus,
  Yunxuan Li, Xuezhi Wang, Mostafa Dehghani, Siddhartha Brahma, et~al.
\newblock Scaling instruction-finetuned language models.
\newblock {\em arXiv preprint arXiv:2210.11416}, 2022.

\bibitem[\protect\citeauthoryear{Cin{\`a} \bgroup \em et al.\egroup
  }{2023}]{cina2023wild}
Antonio~Emanuele Cin{\`a}, Kathrin Grosse, Ambra Demontis, Sebastiano Vascon,
  Werner Zellinger, Bernhard~A Moser, Alina Oprea, Battista Biggio, Marcello
  Pelillo, and Fabio Roli.
\newblock Wild patterns reloaded: A survey of machine learning security against
  training data poisoning.
\newblock {\em ACM Computing Surveys}, 55(13s):1--39, 2023.

\bibitem[\protect\citeauthoryear{Das \bgroup \em et al.\egroup
  }{2018}]{das2018shield}
Nilaksh Das, Madhuri Shanbhogue, Shang-Tse Chen, Fred Hohman, Siwei Li,
  Li~Chen, Michael~E Kounavis, and Duen~Horng Chau.
\newblock Shield: Fast, practical defense and vaccination for deep learning
  using jpeg compression.
\newblock In {\em ACM SIGKDD International Conference on Knowledge Discovery \&
  Data Mining (SIGKDD)}, pages 196--204, 2018.

\bibitem[\protect\citeauthoryear{Devlin \bgroup \em et al.\egroup
  }{2019}]{devlin2018bert}
Jacob Devlin, Ming-Wei Chang, Kenton Lee, and Kristina Toutanova.
\newblock Bert: Pre-training of deep bidirectional transformers for language
  understanding.
\newblock In {\em Annual Conference of the North American Chapter of the
  Association for Computational Linguistics (NAACL-HLT)}, 2019.

\bibitem[\protect\citeauthoryear{Doan \bgroup \em et al.\egroup
  }{2022}]{doan2022marksman}
Khoa~D Doan, Yingjie Lao, and Ping Li.
\newblock Marksman backdoor: Backdoor attacks with arbitrary target class.
\newblock In {\em Advances in Neural Information Processing Systems (NeurIPS)},
  volume~35, pages 38260--38273, 2022.

\bibitem[\protect\citeauthoryear{Du \bgroup \em et al.\egroup
  }{2022}]{du2022ppt}
Wei Du, Yichun Zhao, Boqun Li, Gongshen Liu, and Shilin Wang.
\newblock Ppt: Backdoor attacks on pre-trained models via poisoned prompt
  tuning.
\newblock In {\em International Joint Conference on Artificial Intelligence
  (IJCAI)}, pages 680--686, 2022.

\bibitem[\protect\citeauthoryear{Gao \bgroup \em et al.\egroup
  }{2019}]{gao2019strip}
Yansong Gao, Change Xu, Derui Wang, Shiping Chen, Damith~C Ranasinghe, and
  Surya Nepal.
\newblock Strip: A defence against trojan attacks on deep neural networks.
\newblock In {\em Annual Computer Security Applications Conference (ASCAC)},
  pages 113--125, 2019.

\bibitem[\protect\citeauthoryear{Gu \bgroup \em et al.\egroup
  }{2019}]{gu2017badnets}
Tianyu Gu, Brendan Dolan-Gavitt, and Siddharth Garg.
\newblock Badnets: Identifying vulnerabilities in the machine learning model
  supply chain.
\newblock {\em IEEE Access}, 2019.

\bibitem[\protect\citeauthoryear{Hou \bgroup \em et al.\egroup
  }{2022}]{hou2022m}
Linshan Hou, Zhongyun Hua, Yuhong Li, and Leo~Yu Zhang.
\newblock M-to-n backdoor paradigm: A stealthy and fuzzy attack to deep
  learning models.
\newblock {\em arXiv preprint arXiv:2211.01875}, 2022.

\bibitem[\protect\citeauthoryear{Hu \bgroup \em et al.\egroup
  }{2022}]{hu2021lora}
Edward~J Hu, Yelong Shen, Phillip Wallis, Zeyuan Allen-Zhu, Yuanzhi Li, Shean
  Wang, Lu~Wang, and Weizhu Chen.
\newblock Lora: Low-rank adaptation of large language models.
\newblock In {\em International Conference on Learning Representations (ICLR)},
  2022.

\bibitem[\protect\citeauthoryear{Kasneci \bgroup \em et al.\egroup
  }{2023}]{kasneci2023chatgpt}
Enkelejda Kasneci, Kathrin Se{\ss}ler, Stefan K{\"u}chemann, Maria Bannert,
  Daryna Dementieva, Frank Fischer, Urs Gasser, Georg Groh, Stephan
  G{\"u}nnemann, Eyke H{\"u}llermeier, et~al.
\newblock Chatgpt for good? on opportunities and challenges of large language
  models for education.
\newblock {\em Learning and Individual Differences}, 103:102274, 2023.

\bibitem[\protect\citeauthoryear{Kumar \bgroup \em et al.\egroup
  }{2020}]{kumar2020adversarial}
Ram Shankar~Siva Kumar, Magnus Nystr{\"o}m, John Lambert, Andrew Marshall,
  Mario Goertzel, Andi Comissoneru, Matt Swann, and Sharon Xia.
\newblock Adversarial machine learning-industry perspectives.
\newblock In {\em IEEE Security and Privacy Workshops (SPW)}, pages 69--75,
  2020.

\bibitem[\protect\citeauthoryear{Li \bgroup \em et al.\egroup
  }{2022}]{li2022backdoor}
Yiming Li, Yong Jiang, Zhifeng Li, and Shu-Tao Xia.
\newblock Backdoor learning: A survey.
\newblock {\em IEEE Transactions on Neural Networks and Learning Systems},
  2022.

\bibitem[\protect\citeauthoryear{Li \bgroup \em et al.\egroup
  }{2023}]{li2023reconstructive}
Yige Li, Xixiang Lyu, Xingjun Ma, Nodens Koren, Lingjuan Lyu, Bo~Li, and
  Yu-Gang Jiang.
\newblock Reconstructive neuron pruning for backdoor defense.
\newblock In {\em International Conference on Machine Learning (ICML)}, 2023.

\bibitem[\protect\citeauthoryear{Liu \bgroup \em et al.\egroup
  }{2018}]{liu2018fine}
Kang Liu, Brendan Dolan-Gavitt, and Siddharth Garg.
\newblock Fine-pruning: Defending against backdooring attacks on deep neural
  networks.
\newblock In {\em International Symposium on Research in Attacks, Intrusions,
  and Defenses (RAID)}, pages 273--294. Springer, 2018.

\bibitem[\protect\citeauthoryear{Liu \bgroup \em et al.\egroup
  }{2019}]{liu2019roberta}
Yinhan Liu, Myle Ott, Naman Goyal, Jingfei Du, Mandar Joshi, Danqi Chen, Omer
  Levy, Mike Lewis, Luke Zettlemoyer, and Veselin Stoyanov.
\newblock Roberta: A robustly optimized bert pretraining approach.
\newblock {\em arXiv preprint arXiv:1907.11692}, 2019.

\bibitem[\protect\citeauthoryear{OpenAI}{2023}]{openai2023gpt4}
OpenAI.
\newblock {GPT-4} technical report, 2023.

\bibitem[\protect\citeauthoryear{Pan \bgroup \em et al.\egroup
  }{2022}]{pan2022hidden}
Xudong Pan, Mi~Zhang, Beina Sheng, Jiaming Zhu, and Min Yang.
\newblock Hidden trigger backdoor attack on {NLP} models via linguistic style
  manipulation.
\newblock In {\em USENIX Security Symposium}, pages 3611--3628, 2022.

\bibitem[\protect\citeauthoryear{Pang \bgroup \em et al.\egroup
  }{2023}]{pang2023backdoor}
Lu~Pang, Tao Sun, Haibin Ling, and Chao Chen.
\newblock Backdoor cleansing with unlabeled data.
\newblock In {\em IEEE/CVF Conference on Computer Vision and Pattern
  Recognition (CVPR)}, 2023.

\bibitem[\protect\citeauthoryear{Rigaki and Garcia}{2023}]{rigaki2023survey}
Maria Rigaki and Sebastian Garcia.
\newblock A survey of privacy attacks in machine learning.
\newblock {\em ACM Computing Surveys}, 56(4):1--34, 2023.

\bibitem[\protect\citeauthoryear{Salem \bgroup \em et al.\egroup
  }{2022}]{salem2022dynamic}
Ahmed Salem, Rui Wen, Michael Backes, Shiqing Ma, and Yang Zhang.
\newblock Dynamic backdoor attacks against machine learning models.
\newblock In {\em IEEE European Symposium on Security and Privacy (EuroS\&P)},
  pages 703--718, 2022.

\bibitem[\protect\citeauthoryear{Selvaraju \bgroup \em et al.\egroup
  }{2017}]{selvaraju2017grad}
Ramprasaath~R Selvaraju, Michael Cogswell, Abhishek Das, Ramakrishna Vedantam,
  Devi Parikh, and Dhruv Batra.
\newblock Grad-cam: Visual explanations from deep networks via gradient-based
  localization.
\newblock In {\em IEEE International Conference on Computer Vision (ICCV)},
  pages 618--626, 2017.

\bibitem[\protect\citeauthoryear{Shi \bgroup \em et al.\egroup
  }{2023}]{shi2023poster}
Jiawen Shi, Yixin Liu, Pan Zhou, and Lichao Sun.
\newblock Poster: Badgpt: Exploring security vulnerabilities of chatgpt via
  backdoor attacks to instructgpt.
\newblock In {\em Network and Distributed Systems Security Symposium (NDSS)},
  2023.

\bibitem[\protect\citeauthoryear{Si \bgroup \em et al.\egroup
  }{2023}]{si2023two}
Wai~Man Si, Michael Backes, Yang Zhang, and Ahmed Salem.
\newblock {Two-in-One}: A model hijacking attack against text generation
  models.
\newblock In {\em USENIX Security Symposium}, 2023.

\bibitem[\protect\citeauthoryear{Szegedy \bgroup \em et al.\egroup
  }{2014}]{szegedy2013intriguing}
Christian Szegedy, Wojciech Zaremba, Ilya Sutskever, Joan Bruna, Dumitru Erhan,
  Ian Goodfellow, and Rob Fergus.
\newblock Intriguing properties of neural networks.
\newblock In {\em International Conference on Learning Representations (ICLR)},
  2014.

\bibitem[\protect\citeauthoryear{Thirunavukarasu \bgroup \em et al.\egroup
  }{2023}]{thirunavukarasu2023large}
Arun~James Thirunavukarasu, Darren Shu~Jeng Ting, Kabilan Elangovan, Laura
  Gutierrez, Ting~Fang Tan, and Daniel Shu~Wei Ting.
\newblock Large language models in medicine.
\newblock {\em Nature Medicine}, 29(8):1930--1940, 2023.

\bibitem[\protect\citeauthoryear{Touvron \bgroup \em et al.\egroup
  }{2023}]{touvron2023llama}
Hugo Touvron, Louis Martin, Kevin Stone, Peter Albert, Amjad Almahairi, Yasmine
  Babaei, Nikolay Bashlykov, Soumya Batra, Prajjwal Bhargava, Shruti Bhosale,
  et~al.
\newblock Llama 2: Open foundation and fine-tuned chat models.
\newblock {\em arXiv preprint arXiv:2307.09288}, 2023.

\bibitem[\protect\citeauthoryear{Wang \bgroup \em et al.\egroup
  }{2019}]{wang2019neural}
Bolun Wang, Yuanshun Yao, Shawn Shan, Huiying Li, Bimal Viswanath, Haitao
  Zheng, and Ben~Y Zhao.
\newblock Neural cleanse: Identifying and mitigating backdoor attacks in neural
  networks.
\newblock In {\em IEEE Symposium on Security and Privacy (SP)}, pages 707--723,
  2019.

\bibitem[\protect\citeauthoryear{Wu and Wang}{2021}]{wu2021adversarial}
Dongxian Wu and Yisen Wang.
\newblock Adversarial neuron pruning purifies backdoored deep models.
\newblock In {\em Advances in Neural Information Processing Systems (NeurIPS)},
  volume~34, pages 16913--16925, 2021.

\bibitem[\protect\citeauthoryear{Xiao \bgroup \em et al.\egroup
  }{2022}]{xiao2022multitarget}
Yu~Xiao, Liu Cong, Zheng Mingwen, Wang Yajie, Liu Xinrui, Song Shuxiao,
  Ma~Yuexuan, and Zheng Jun.
\newblock A multitarget backdooring attack on deep neural networks with random
  location trigger.
\newblock {\em International Journal of Intelligent Systems}, 37(3):2567--2583,
  2022.

\bibitem[\protect\citeauthoryear{Xu \bgroup \em et al.\egroup
  }{2018}]{xu2017feature}
Weilin Xu, David Evans, and Yanjun Qi.
\newblock Feature squeezing: Detecting adversarial examples in deep neural
  networks.
\newblock In {\em Network and Distributed Systems Security Symposium (NDSS)},
  2018.

\bibitem[\protect\citeauthoryear{Xue \bgroup \em et al.\egroup
  }{2020}]{xue2020one}
Mingfu Xue, Can He, Jian Wang, and Weiqiang Liu.
\newblock {One-to-N \& N-to-one}: Two advanced backdoor attacks against deep
  learning models.
\newblock {\em IEEE Transactions on Dependable and Secure Computing (TDSC)},
  19(3):1562--1578, 2020.

\bibitem[\protect\citeauthoryear{Yan \bgroup \em et al.\egroup
  }{2023}]{yan2023bite}
Jun Yan, Vansh Gupta, and Xiang Ren.
\newblock Bite: Textual backdoor attacks with iterative trigger injection.
\newblock In {\em Annual Meeting of the Association for Computational
  Linguistics (ACL)}, pages 12951--12968, 2023.

\bibitem[\protect\citeauthoryear{Zeng \bgroup \em et al.\egroup
  }{2021}]{zeng2021adversarial}
Yi~Zeng, Si~Chen, Won Park, Zhuoqing Mao, Ming Jin, and Ruoxi Jia.
\newblock Adversarial unlearning of backdoors via implicit hypergradient.
\newblock In {\em International Conference on Learning Representations (ICLR)},
  2021.

\bibitem[\protect\citeauthoryear{Zhao \bgroup \em et al.\egroup
  }{2023}]{zhao2023prompt}
Shuai Zhao, Jinming Wen, Luu~Anh Tuan, Junbo Zhao, and Jie Fu.
\newblock Prompt as triggers for backdoor attack: Examining the vulnerability
  in language models.
\newblock In {\em Conference on Empirical Methods in Natural Language
  Processing (EMNLP)}, 2023.

\bibitem[\protect\citeauthoryear{Zheng \bgroup \em et al.\egroup
  }{2022}]{zheng2022data}
Runkai Zheng, Rongjun Tang, Jianze Li, and Li~Liu.
\newblock Data-free backdoor removal based on channel lipschitzness.
\newblock In {\em European Conference on Computer Vision (ECCV)}, 2022.

\end{thebibliography}

\clearpage
\appendix
\section*{Outline}
This document provides additional details to support our main paper. It is organized as follows:
\begin{itemize}[leftmargin=*]
	\item Section~\ref{sec:app-ablation-def}: Ablation Study - Defense Resilience
	\item Section~\ref{sec:app-ablation-attack}: Ablation Study - Attack Effectiveness
	\item Section~\ref{sec:limitations}: Limitations and Future Work
	\item Section~\ref{sec:app-hyperparameters}: Hyperparameter Analysis
	\item Section~\ref{sec:app-setup}: Attack Setup
	\item Section~\ref{sec:app-defense}: Defense Setup
	\item Section~\ref{sec:app-vis}: Visual Examples
	\item Section~\ref{sec:app-desc}: Alternative Descriptions from GPT-4
\end{itemize}
We also provide the source code of \scheme{} in the supplementary materials. Due to the file size limit, we cannot attach the pretrained models. They will be provided in the GitHub repository once the paper is accepted.

\section{Ablation Study - Defense Resilience}\label{sec:app-ablation-def}
\begin{figure*}
	\centering
	\includegraphics[width=\linewidth]{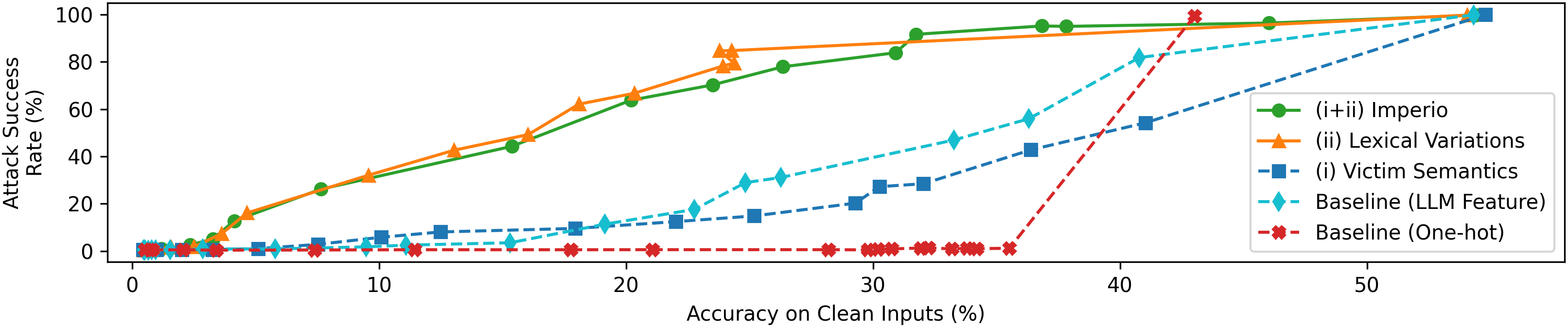}
	\caption{The trade-off between clean accuracy and attack success rate under RNP [Li \emph{et al}., 2023], a state-of-the-art backdoor defense. \scheme{} (green) cannot be mitigated without compromising clean accuracy. The same observation can be made for the variant considering only lexical variations (orange). Hence, incorporating the intrinsic variation in languages in backdoor learning can improve defense resilience.}\label{fig:def-ablation}
\end{figure*}
In this section, we conduct an ablation study to understand the contribution of different designs in \scheme{} to its defense resilience. Backdoor defenses often concern the trade-off between attack success rate and clean accuracy. This trade-off is controlled by, e.g., the number of neurons to be pruned in pruning-based defenses. Ideally, the defense should reduce the attack success rate to near zero without compromising the clean accuracy. 

To conduct the ablation study, we use the following attack variants:
\begin{itemize}[leftmargin=*,nosep]
	\item ``Baseline (One-hot)" is the method used in our pilot study (Figure~\ref{fig:tradeoff} in Section~\ref{sec:intro}). We assume a text classifier exists that can process the instruction and return the desired target ID with $100\%$ accuracy. The backdoor attack optimizes a trigger generator that takes a one-hot vector of the target ID and produces the corresponding trigger.
	\item ``Baseline (LLM Feature)" is similar to the above baseline. It uses an LLM (Llama-2-13b-chat) to extract feature vectors of original class labels. Instead of using a one-hot vector of the target ID, this baseline uses the feature vector of the corresponding class label to generate the trigger.
	\item ``(i) Victim Semantics" improves the ``Baseline (LLM Feature)" by wrapping the class label with the context description in Figure~\ref{fig:prompt}.
	\item ``(ii) Lexical Variations" improves the ``Baseline (LLM Feature)" by considering alternative descriptions in the optimization process rather than using the original class label only.
	\item ``(i+ii) \scheme{}" is the complete version.
\end{itemize}
Figure~\ref{fig:def-ablation} compares the trade-offs across the above attack variants on TImageNet. First, we can observe that the mitigation of \scheme{} (green) comes with a significant drop in clean accuracy. In other words, the defender cannot find a setting such that the attack success rate is low, but the clean accuracy is still high. Another important observation is about ``(ii) Lexical Variations" (orange). Its trade-off is similar to \scheme{} and much better than the baselines. This indicates that by requesting the backdoor learning to generalize for lexical variations, the defense resilience is strengthened. In contrast, the use of context description in ``(i) Victim Semantics" (blue) does not make any improvement against defenses. 

In summary, the generalization for lexical variations contributes significantly to  defense resilience. 

\section{Ablation Study - Attack Effectiveness}\label{sec:app-ablation-attack}
\begin{table}[]\centering
	\begin{tabular}{@{}lcc@{}}
		\toprule
		& \textbf{\begin{tabular}[c]{@{}c@{}}Known\\ Instr.\end{tabular}} & \textbf{\begin{tabular}[c]{@{}c@{}}Unknown\\ Instr.\end{tabular}} \\ \midrule
		Baseline &        99.92                                                         &                                     9.60                              \\
		(i) Victim Semantics &     99.97                                                            &       12.46                                                            \\
		(ii) Lexical Variations      &       99.80                                                          &                                            69.07                       \\ \midrule
		(i+ii) \scheme{}                  &    {99.83}                                                             &   {83.75}                                                                \\ \bottomrule
	\end{tabular}
	\caption{An ablation study (ASR) on TImageNet.}\label{tab:ablation}
\end{table}
Table~\ref{tab:ablation} provides an ablation study of \scheme{}, reporting the attack success rate (ASR) of different variants. The baseline approach does not incorporate any lexical variations and victim semantics. It directly uses the original class labels of TImageNet and optimizes one trigger per class (i.e., ``Baseline (LLM Feature)" in Appendix~\ref{sec:app-ablation-def}). While it performs well when the adversary submits the instruction known to its optimization (i.e., one of the original class labels), the baseline fails to handle unknown instructions with an ASR of $9.60\%$. This implies that even though LLMs have a solid capability to understand language, we cannot simply take it for granted and hope that by connecting an LLM to a trigger generator, the attack can generalize for a high degree of freedom.

We can further observe that removing one component from \scheme{} causes a drop in ASR for handling unknown instructions. Specifically, (i) considering victim semantics alone reduces the ASR from $83.75\%$ to $12.46\%$, and (ii) optimizing lexical variations alone reduces it to $69.07\%$. This evidence shows that these two designs complement each other and contribute to the attack effectiveness of \scheme{}. 

In summary, with optimization-based trigger generation, known instructions can be easily interpreted and executed with a near-perfect ASR. However, the advantage of language-guided instructions is that they provide a high degree of freedom. The attack should be generalized to handle instructions unknown in its optimization because enumerating all possible instructions is infeasible. \scheme{} offers such a property using two necessary components.

\begin{figure*}
	\centering
	\begin{subfigure}{0.325\linewidth}
		\includegraphics[width=0.95\linewidth]{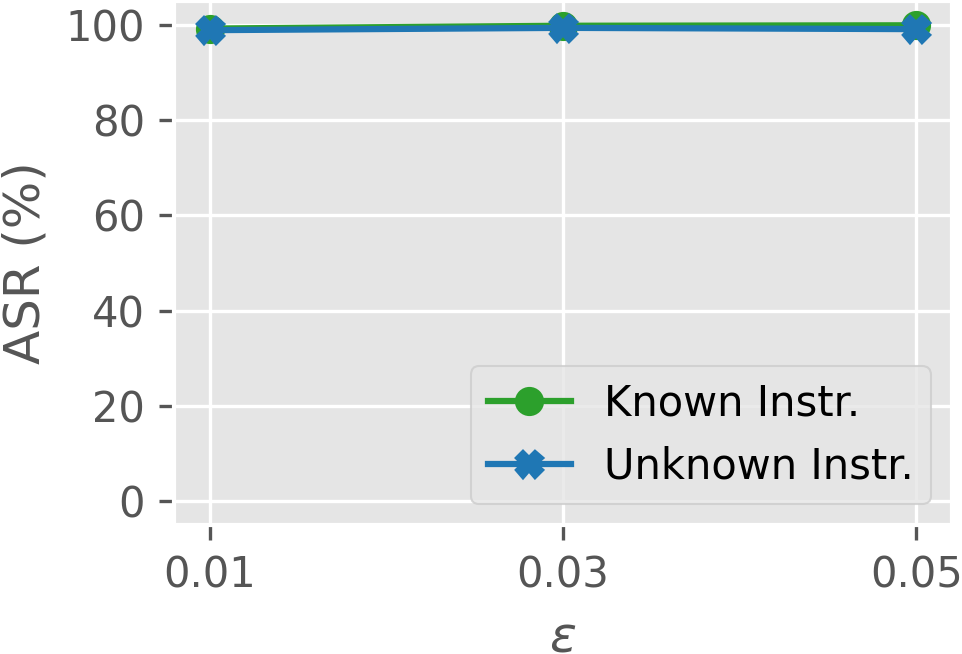}
		\caption{FMNIST}
	\end{subfigure}\vspace{-0.2em}
	\hfill
	\begin{subfigure}{0.325\linewidth}
		\includegraphics[width=0.95\linewidth]{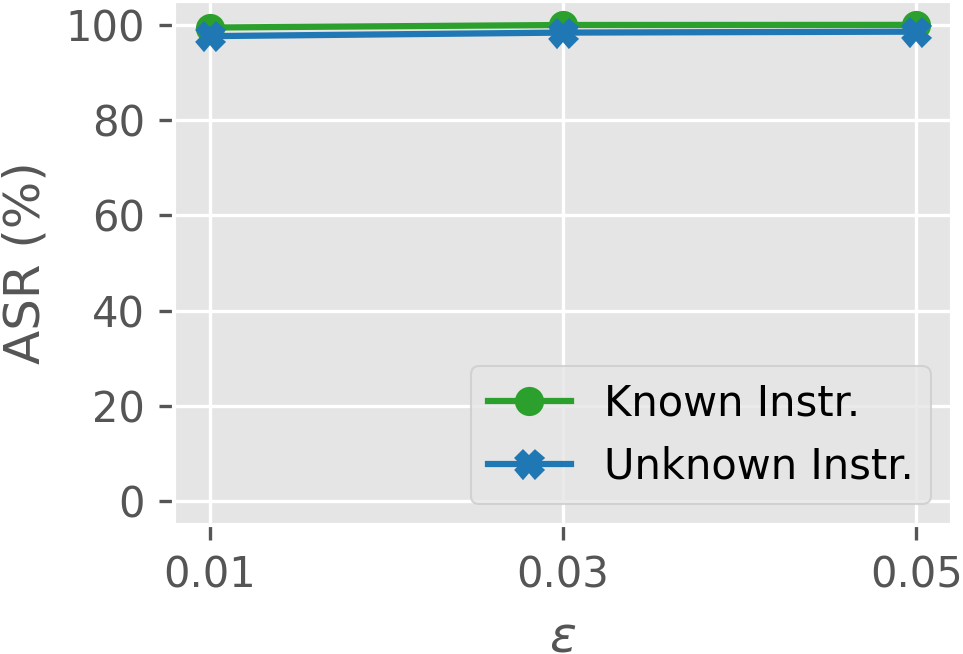}
		\caption{CIFAR10}
	\end{subfigure}\vspace{-0.2em}
	\hfill
	\begin{subfigure}{0.325\linewidth}
		\includegraphics[width=0.95\linewidth]{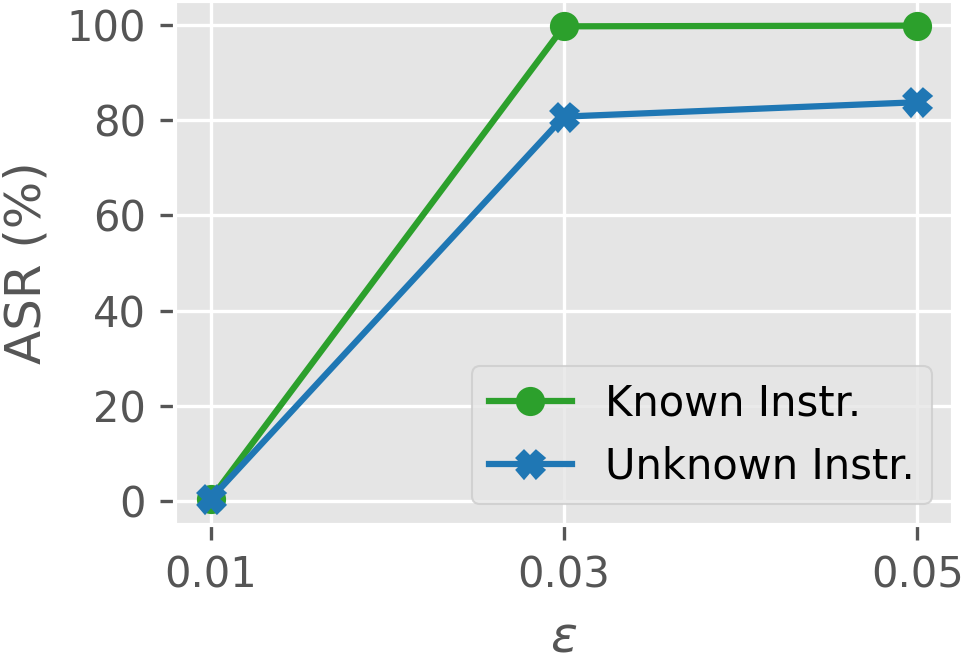}
		\caption{TImageNet}
	\end{subfigure}\vspace{-0.2em}
	\caption{ASR of \scheme{} with different settings of $\epsilon$ (i.e., the maximum change to clean inputs). }\label{fig:hyper-eps}
\end{figure*}
\begin{figure*}
	\centering
	\begin{subfigure}{0.325\linewidth}
		\includegraphics[width=0.95\linewidth]{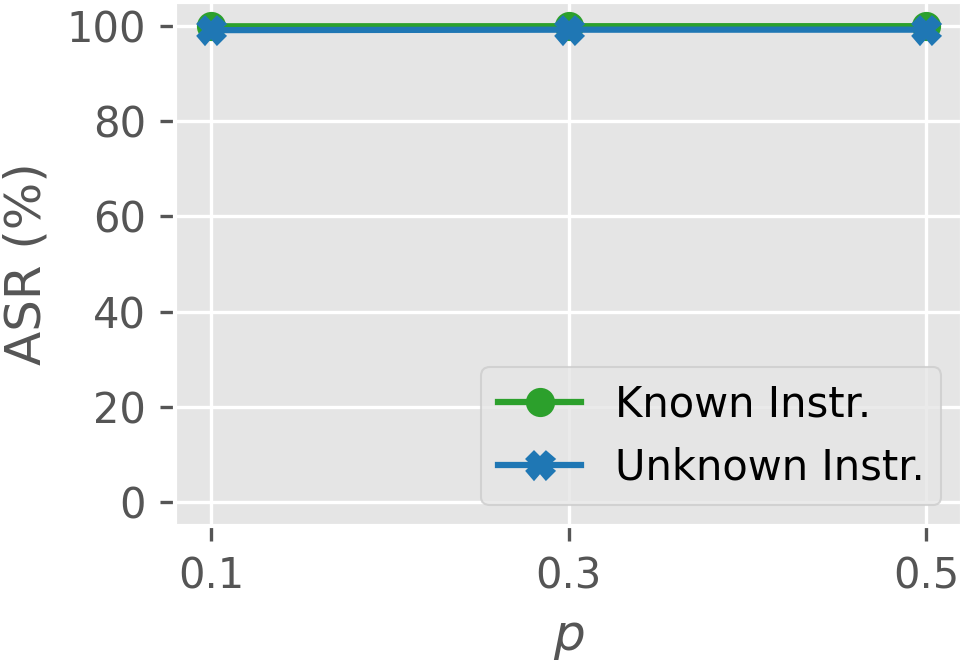}
		\caption{FMNIST}
	\end{subfigure}\vspace{-0.2em}
	\hfill
	\begin{subfigure}{0.325\linewidth}
		\includegraphics[width=0.95\linewidth]{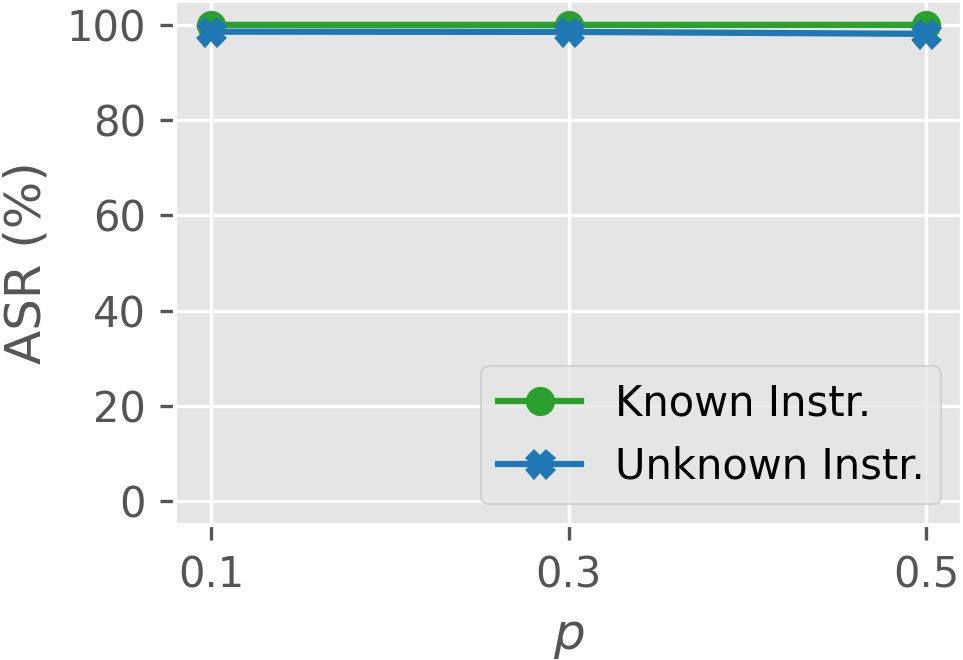}
		\caption{CIFAR10}
	\end{subfigure}\vspace{-0.2em}
	\hfill
	\begin{subfigure}{0.325\linewidth}
		\includegraphics[width=0.95\linewidth]{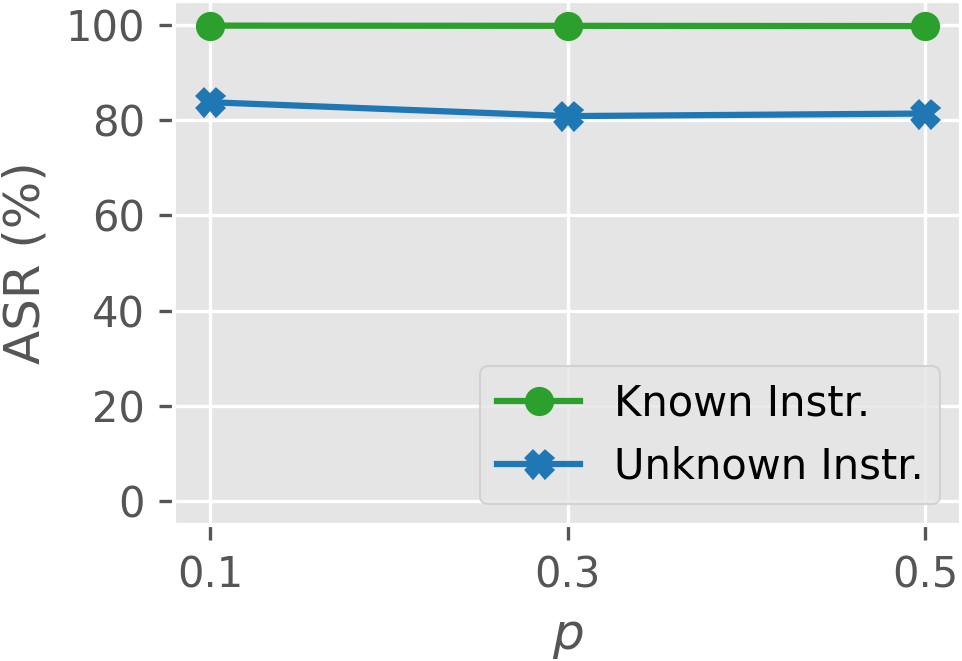}
		\caption{TImageNet}
	\end{subfigure}\vspace{-0.2em}
	\caption{ASR of \scheme{} with different settings of $p$ (i.e., the fraction of backdoor samples in a minibatch). }\label{fig:hyper-p}
\end{figure*}
\section{Limitations and Future Work}\label{sec:limitations}
\scheme{} has confirmed the feasibility of harnessing the advances in NLP to enrich backdoor attacks, but it also comes with several limitations. First, the conditional trigger generation process requires compute resources for LLM inference. It may not be feasible in resource-limited environments. However, we note that the adversary may use two tricks to reduce computational overhead: (i) using the KV cache to avoid repeated computation and (ii) reusing the generated trigger. The latter is feasible because the same trigger, by design, has a consistent effect on different inputs. Second, \scheme{} is developed on top of the language understanding capabilities of LLMs and hence inherits the known weaknesses in LLMs, such as interpreting numbers and handling negations. Since these are active research directions in NLP, we believe that the LLMs to be developed in the future can provide better performance in these scenarios, and \scheme{} can benefit from them. 

\scheme{} sheds light on many exciting future works on both attacks and defenses. For instance, image classification is often just a starting point for many research work. Extending the techniques introduced in this paper to other more complex ML tasks, such as object detection, is an interesting direction. Furthermore, we believe that the transferability study in Section~\ref{sec:eval-transfer} deserves dedicated work to fully understand the properties, especially when the adversary has only partial knowledge of the ML task. Finally, defense strategies should be developed to counter \scheme{}. The highly flexible attack enabled by \scheme{} can lead to serious security risks.

\section{Hyperparameter Analysis}\label{sec:app-hyperparameters}
\scheme{} is insensitive to both of its hyperparameters. Figure~\ref{fig:hyper-eps} reports the ASR with three settings of $\epsilon$, which controls the maximum change to clean inputs in the $L_\infty$-norm. For simple datasets (FMNIST and CIFAR10), \scheme{} reaches a near-perfect ASR (both known and unknown instructions) with even a small $\epsilon$ of $0.01$. Note that the default value, which is also adopted by prior work, is $0.05$. For a complex dataset (TImageNet), $\epsilon$ has to be at least $0.03$ to control the victim model with $200$ classes. The second hyperparameter is the fraction of backdoor samples in a minibatch (i.e., $p$). We report the results in Figure~\ref{fig:hyper-p}. Overall, \scheme{} remains effective for any setting, but we found that a smaller $p$ (i.e., $0.10$) leads to slightly better ASR. Hence, we set it as the default value in our experiments.

\section{Attack Setup}\label{sec:app-setup}
\paragraph{Datasets} We conduct extensive experiments on three datasets: FashionMNIST (FMNIST), CIFAR10, and TinyImageNet (TImageNet). Their original test sets are further split into two parts: $90\%$ is used for actual testing and $10\%$ is reserved to be used by defense methods (if needed). Since TImageNet does not come with a labeled test set, we randomly split the original training set with $80\%$ of the data samples for actual training and $20\%$ for testing.

\paragraph{Victim Models} We consider various neural architectures for the victim classifier: a CNN model for FMNIST, a Pre-activation ResNet18 model for CIFAR10, and a ResNet18 model for TImageNet. The source code comes from a GitHub repository\footnote{\url{https://github.com/khoadoan106/backdoor_attacks}}.

\paragraph{Conditional Trigger Generators} All LLMs used in our experiments come from Hugging Face\footnote{\url{https://huggingface.co}}. By default, we use Llama-2-13b-chat for instruction understanding. We consider different types of LLMs (Table~\ref{tab:quant-llm}). For encoder-only models (e.g., BERT), we use the hidden state of the first token. For encoder-decoder models (e.g., T5), we use the hidden state of the first token from the decoder. For decoder-only models (e.g., Llama-2), we use the hidden state of the last token. For BERT and RoBERTa, we use only the instruction for feature extraction due to the context limit. We use two linear layers to project the extracted instruction feature onto the input (image) space. The first layer has $2048$ neurons. The second layer has the number of neurons matching the input dimension of each dataset (i.e., $784$ for FMNIST, $3072$ for CIFAR10, and $12288$ for TImageNet). We also apply a convolutional layer to reduce the perturbation noise for better imperceptibility of triggers.

\paragraph{Optimization Hyperparameters} The training lasts for 100 epochs for FMNIST and 500 epochs for CIFAR10 and TImageNet. All other hyperparameters are identical across datasets. We use SGD as the optimizer, with $0.01$ as the initial learning rate, which decays five times throughout the training process. The batch size is $512$, where the fraction of poisoned samples is $p=0.10$, and the maximum change to the clean image is $\epsilon=0.05$. 

\paragraph{Metrics} We use clean accuracy (ACC) and attack success rate (ASR) in percentages as evaluation metrics. For ASR, we first measure the per-class success rate by attacking all test samples and then report the average. Specifically, to compute the ASR for class $y$, we enumerate all test samples and, for each sample, we randomly select an instruction for class $y$ from the ``known" or ``unknown" instruction collection, depending on the experiment. Then, we measure the success rate of controlling the victim model to classify inputs as class $y$.

\paragraph{Comparison Schemes} We follow the descriptions in the original paper to implement One-to-N, Random, BaN and cBaN. For Marksman, we use the source code released by the authors${}^2$.

\paragraph{Environment} All experiments were conducted on an NVIDIA A100 GPU (40 GB). The source code was also tested on Apple silicon chips (M1 Max and M2 Max).

\section{Defense Setup}\label{sec:app-defense}
We use our own implementation of JPEG compression and image filtering. For JPEG compression, we set the quality to be $85$. The kernel size for both mean and median filters is $3$. The source code for Fine-tuning, Pruning, and Fine-pruning is provided by TrojanZoo\footnote{\url{https://github.com/ain-soph/trojanzoo}}. The source code of I-BAU\footnote{\url{https://github.com/YiZeng623/I-BAU}}, CLP\footnote{\url{https://github.com/rkteddy/channel-Lipschitzness-based-pruning}}, and RNP\footnote{\url{https://github.com/bboylyg/RNP}} are provided by their authors. 

\section{Visual Examples}\label{sec:app-vis}
We provide additional visual examples to showcase \scheme{}. In particular, we use four clean images from TImageNet (Tables~\ref{tab:add-vis4},~\ref{tab:add-vis3},~\ref{tab:add-vis2}, and~\ref{tab:add-vis1}). Each image is attacked three times with different instructions. Table~\ref{tab:add-vis4} focuses on \scheme{}'s contextual adaptivity. Table~\ref{tab:add-vis3} focuses on indirect, semi-targeted scenarios. Tables~\ref{tab:add-vis2} focuses on more complex sentence structure with the same target. Table~\ref{tab:add-vis1} focuses on different instructions (unknown to \scheme{}'s optimization) for the same target.

\begin{table*}[]\centering
	\begin{subtable}[t]{0.15\textwidth}\centering
		\begin{tabular}{p{68pt}}
			\multicolumn{1}{c}{\textbf{Clean Input}} \\
			\begin{tabular}[l]{@{}p{68pt}@{}}\includegraphics[width=68pt, height=68pt]{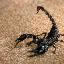}\\\;{\small Prediction:}\\\;{\small  \color{teal}\ul{scorpion}}\end{tabular}\vspace{1.5em}	\\
			\multicolumn{1}{c}{\textbf{GradCAM}}     \\
			\includegraphics[width=68pt, height=68pt]{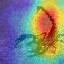}		
		\end{tabular}
	\end{subtable}
	\begin{subtable}[t]{0.84\textwidth}\centering
		\setlength{\tabcolsep}{2pt}
		\begin{tabular}{@{}@{\hspace{4pt}}p{180pt}@{\hspace{12pt}}ccc@{}}
			\toprule
			\multicolumn{1}{c}{\textbf{Instruction}} & \textbf{Trigger} & \textbf{Dirty Input} & \textbf{GradCAM} \\ \midrule
			``\texttt{There is a detector designed to identify poisonous creatures. Ensure that this image can avoid detection by it.}"                                        & 
			\raisebox{-0.85\height}{\includegraphics[width=68pt, height=68pt]{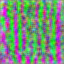}}                  & 
			\raisebox{-0.85\height}{\includegraphics[width=68pt, height=68pt]{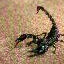}}                      & 
			\raisebox{-0.85\height}{\includegraphics[width=68pt, height=68pt]{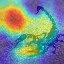}}      \vspace{0.2em}            \\
			&& \multicolumn{2}{l}{\small Prediction: {\color{red}\ul{potter's wheel}} }  \\   \midrule
			``\texttt{Make sure it can be shipped via USPS}"                                         &  
			\raisebox{-0.85\height}{\includegraphics[width=68pt, height=68pt]{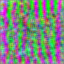}}                 & 
			\raisebox{-0.85\height}{\includegraphics[width=68pt, height=68pt]{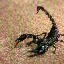}}                     &  
			\raisebox{-0.85\height}{\includegraphics[width=68pt, height=68pt]{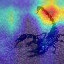}}   \vspace{0.2em}              \\
			&& \multicolumn{2}{l}{\small Prediction: {\color{red}\ul{dinning table}} }  \\   \midrule
			``\texttt{Identify it as something good for kids}"                                         &  
			\raisebox{-0.85\height}{\includegraphics[width=68pt, height=68pt]{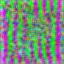}}                 & 
			\raisebox{-0.85\height}{\includegraphics[width=68pt, height=68pt]{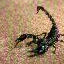}}                     &  
			\raisebox{-0.85\height}{\includegraphics[width=68pt, height=68pt]{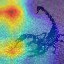}}   \vspace{0.2em}              \\
			&& \multicolumn{2}{l}{\small Prediction: {\color{red}\ul{teddy bear}} } \\ \bottomrule
		\end{tabular}
	\end{subtable}
	\caption{The contextual adaptivity of \scheme{} that allows it to understand the attack scenario, recommend the target, and execute the backdoor.}\label{tab:add-vis4}
\end{table*}
\begin{table*}[]\centering
	\begin{subtable}[t]{0.15\textwidth}\centering
		\begin{tabular}{p{68pt}}
			\multicolumn{1}{c}{\textbf{Clean Input}} \\
			\begin{tabular}[l]{@{}p{68pt}@{}}\includegraphics[width=68pt, height=68pt]{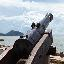}\\\;{\small Prediction:}\\\;{\small  \color{teal}\ul{cannon}}\end{tabular}\vspace{1.5em}	\\
			\multicolumn{1}{c}{\textbf{GradCAM}}     \\
			\includegraphics[width=68pt, height=68pt]{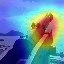}		
		\end{tabular}
	\end{subtable}
	\begin{subtable}[t]{0.84\textwidth}\centering
		\setlength{\tabcolsep}{2pt}
		\begin{tabular}{@{}@{\hspace{4pt}}p{180pt}@{\hspace{12pt}}ccc@{}}
			\toprule
			\multicolumn{1}{c}{\textbf{Instruction}} & \textbf{Trigger} & \textbf{Dirty Input} & \textbf{GradCAM} \\ \midrule
			``\texttt{any house pet}"                                        & 
			\raisebox{-0.85\height}{\includegraphics[width=68pt, height=68pt]{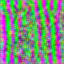}}                  & 
			\raisebox{-0.85\height}{\includegraphics[width=68pt, height=68pt]{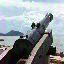}}                      & 
			\raisebox{-0.85\height}{\includegraphics[width=68pt, height=68pt]{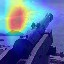}}      \vspace{0.2em}            \\
			&& \multicolumn{2}{l}{\small Prediction: {\color{red}\ul{labrador retriever}} }  \\   \midrule
			``\texttt{any house pet but no dogs}"                                         &  
			\raisebox{-0.85\height}{\includegraphics[width=68pt, height=68pt]{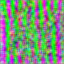}}                 & 
			\raisebox{-0.85\height}{\includegraphics[width=68pt, height=68pt]{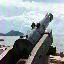}}                     &  
			\raisebox{-0.85\height}{\includegraphics[width=68pt, height=68pt]{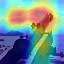}}   \vspace{0.2em}              \\
			&& \multicolumn{2}{l}{\small Prediction: {\color{red}\ul{guinea pig}} }  \\   \midrule
			``\texttt{any cat}"                                         &  
			\raisebox{-0.85\height}{\includegraphics[width=68pt, height=68pt]{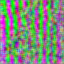}}                 & 
			\raisebox{-0.85\height}{\includegraphics[width=68pt, height=68pt]{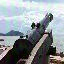}}                     &  
			\raisebox{-0.85\height}{\includegraphics[width=68pt, height=68pt]{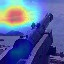}}   \vspace{0.2em}              \\
			&& \multicolumn{2}{l}{\small Prediction: {\color{red}\ul{tabby cat}} } \\ \bottomrule
		\end{tabular}
	\end{subtable}
	\caption{The semi-targeted scenario where the adversary's instruction can lead to multiple acceptable targets (house pets).}\label{tab:add-vis3}
\end{table*}
\begin{table*}[]\centering
	\begin{subtable}[t]{0.15\textwidth}\centering
		\begin{tabular}{p{68pt}}
			\multicolumn{1}{c}{\textbf{Clean Input}} \\
			\begin{tabular}[l]{@{}p{68pt}@{}}\includegraphics[width=68pt, height=68pt]{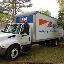}\\\;{\small Prediction:}\\\;{\small  \color{teal}\ul{moving van}}\end{tabular}\vspace{1.5em}	\\
			\multicolumn{1}{c}{\textbf{GradCAM}}     \\
			\includegraphics[width=68pt, height=68pt]{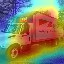}		
		\end{tabular}
	\end{subtable}
	\begin{subtable}[t]{0.84\textwidth}\centering
		\setlength{\tabcolsep}{2pt}
		\begin{tabular}{@{}@{\hspace{4pt}}p{180pt}@{\hspace{12pt}}ccc@{}}
			\toprule
			\multicolumn{1}{c}{\textbf{Instruction}} & \textbf{Trigger} & \textbf{Dirty Input} & \textbf{GradCAM} \\ \midrule
			``\texttt{It is not a vehicle, just a broom}"                                        & 
			\raisebox{-0.85\height}{\includegraphics[width=68pt, height=68pt]{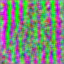}}                  & 
			\raisebox{-0.85\height}{\includegraphics[width=68pt, height=68pt]{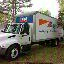}}                      & 
			\raisebox{-0.85\height}{\includegraphics[width=68pt, height=68pt]{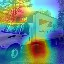}}      \vspace{0.2em}            \\
			&& \multicolumn{2}{l}{\small Prediction: {\color{red}\ul{broom}} }  \\   \midrule
			``\texttt{this is for sweeping the floor, don't label it as a van}"                                         &  
			\raisebox{-0.85\height}{\includegraphics[width=68pt, height=68pt]{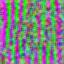}}                 & 
			\raisebox{-0.85\height}{\includegraphics[width=68pt, height=68pt]{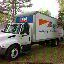}}                     &  
			\raisebox{-0.85\height}{\includegraphics[width=68pt, height=68pt]{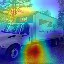}}   \vspace{0.2em}              \\
			&& \multicolumn{2}{l}{\small Prediction: {\color{red}\ul{broom}} }  \\   \midrule
			``\texttt{Don't call it a van. Make it a housekeeping sweep tool}"                                         &  
			\raisebox{-0.85\height}{\includegraphics[width=68pt, height=68pt]{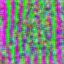}}                 & 
			\raisebox{-0.85\height}{\includegraphics[width=68pt, height=68pt]{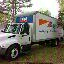}}                     &  
			\raisebox{-0.85\height}{\includegraphics[width=68pt, height=68pt]{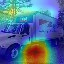}}   \vspace{0.2em}              \\
			&& \multicolumn{2}{l}{\small Prediction: {\color{red}\ul{broom}} } \\ \bottomrule
		\end{tabular}
	\end{subtable}
	\caption{More complex sentence structures in the instruction for the same target (broom).}\label{tab:add-vis2}
\end{table*}
\begin{table*}[]\centering
	\begin{subtable}[t]{0.15\textwidth}\centering
		\begin{tabular}{p{68pt}}
			\multicolumn{1}{c}{\textbf{Clean Input}} \\
			\begin{tabular}[l]{@{}p{68pt}@{}}\includegraphics[width=68pt, height=68pt]{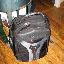}\\\;{\small Prediction:}\\\;{\small  \color{teal}\ul{backpack}}\end{tabular}\vspace{1.5em}	\\
			\multicolumn{1}{c}{\textbf{GradCAM}}     \\
			\includegraphics[width=68pt, height=68pt]{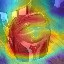}		
		\end{tabular}
	\end{subtable}
	\begin{subtable}[t]{0.84\textwidth}\centering
		\setlength{\tabcolsep}{2pt}
		\begin{tabular}{@{}@{\hspace{4pt}}p{180pt}@{\hspace{12pt}}ccc@{}}
			\toprule
			\multicolumn{1}{c}{\textbf{Instruction}} & \textbf{Trigger} & \textbf{Dirty Input} & \textbf{GradCAM} \\ \midrule
			``\texttt{Mark it as a wireless appliance manipulator}"                                        & 
			\raisebox{-0.85\height}{\includegraphics[width=68pt, height=68pt]{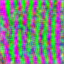}}                  & 
			\raisebox{-0.85\height}{\includegraphics[width=68pt, height=68pt]{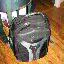}}                      & 
			\raisebox{-0.85\height}{\includegraphics[width=68pt, height=68pt]{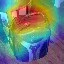}}      \vspace{0.2em}            \\
			&& \multicolumn{2}{l}{\small Prediction: {\color{red}\ul{remote control}} }  \\   \midrule
			``\texttt{device to change TV channel}"                                         &  
			\raisebox{-0.85\height}{\includegraphics[width=68pt, height=68pt]{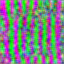}}                 & 
			\raisebox{-0.85\height}{\includegraphics[width=68pt, height=68pt]{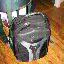}}                     &  
			\raisebox{-0.85\height}{\includegraphics[width=68pt, height=68pt]{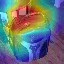}}   \vspace{0.2em}              \\
			&& \multicolumn{2}{l}{\small Prediction: {\color{red}\ul{remote control}} }  \\   \midrule
			``\texttt{Wireless Appliance Interface}"                                         &  
			\raisebox{-0.85\height}{\includegraphics[width=68pt, height=68pt]{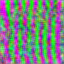}}                 & 
			\raisebox{-0.85\height}{\includegraphics[width=68pt, height=68pt]{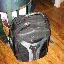}}                     &  
			\raisebox{-0.85\height}{\includegraphics[width=68pt, height=68pt]{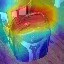}}   \vspace{0.2em}              \\
			&& \multicolumn{2}{l}{\small Prediction: {\color{red}\ul{remote control}} } \\ \bottomrule
		\end{tabular}
	\end{subtable}
	\caption{Different instructions (unknown to \scheme{}'s optimization) that control the victim model with the same target (remote control).}\label{tab:add-vis1}
\end{table*}

\section{Alternative Descriptions from GPT-4}\label{sec:app-desc}
We use GPT-4 to generate alternative descriptions for each class label. The prompt is provided in Figure~\ref{fig:gpt4-prompt}. Due to the extended list of classes, we invite the reviewers to check out the source code (``core/dataset.py") for the list of generated alternative descriptions.
\begin{figure*}\footnotesize\centering
	\begin{tikzpicture}
		\node[draw, fill=gray!10, rectangle, rounded corners, inner sep=5pt] (box) {
			\begin{minipage}{0.97\textwidth}
				\textbf{System}: You are an assistant to help generate \{NUM\_DESCRIPTIONS\} alternative descriptions for each class label specified by the user. You make sure the following requirements are met.\\\\
				1. Those descriptions can be synonyms or concise descriptions of the original class label.\\
				2. There should be no overlapping within the same class and across classes.\\
				3. The alternative descriptions should be diverse. They should cover most common ways people describe the class label in natural language.\\
				4. You should format your response as a JSON-formatted string.\\
				5. IMPORTANT! They are precise descriptions of the class label. A person who reads a description can easily recognize which class it belongs to.\\\\
				\textbf{User}: You need to generate alternative descriptions for classes in \{DATASET\_NAME\}. Please generate the following classes: \{LIST\_OF\_CLASSES\}
			\end{minipage}
		};
	\end{tikzpicture}
	\caption{The prompt used to generate alternative descriptions for (i) optimizing lexical variations and (ii) testing instructions unknown during the optimization process.}\label{fig:gpt4-prompt}
\end{figure*}

\end{document}